\documentclass[a4paper,fleqn,usenatbib]{mnras}

\usepackage{newtxtext,newtxmath}

\usepackage[T1]{fontenc}
\usepackage{ae,aecompl}

\usepackage{graphicx}	
\usepackage{amsmath}	
\usepackage{amssymb}	
\usepackage{amsbsy}
\usepackage{algorithm}
\usepackage{algorithmic}
\usepackage{bbm}
\usepackage{array}
\usepackage{mathrsfs}
\usepackage{float}
\usepackage{mathrsfs}
\usepackage[utf8]{inputenc}
\usepackage{mathtools}
\usepackage[dvipsnames,table]{xcolor}
\usepackage{lipsum}
\usepackage{pifont}
\usepackage{chngcntr}
\usepackage{hyperref}
\hypersetup{
    colorlinks = true,
    urlcolor = black,
    linkbordercolor = {white}
}
\usepackage[most]{tcolorbox}
\usepackage{etoolbox}
\usepackage{longtable}
\usepackage{pdflscape}
\usepackage{geometry}
\usepackage{enumitem}
\usepackage{booktabs}
\usepackage{multirow}

\makeatletter

\patchcmd{\NAT@citex}
  {\@citea\NAT@hyper@{%
     \NAT@nmfmt{\NAT@nm}%
     \hyper@natlinkbreak{\NAT@aysep\NAT@spacechar}{\@citeb\@extra@b@citeb}%
     \NAT@date}}
  {\@citea\NAT@nmfmt{\NAT@nm}%
   \NAT@aysep\NAT@spacechar\NAT@hyper@{\NAT@date}}{}{}

\patchcmd{\NAT@citex}
  {\@citea\NAT@hyper@{%
     \NAT@nmfmt{\NAT@nm}%
     \hyper@natlinkbreak{\NAT@spacechar\NAT@@open\if*#1*\else#1\NAT@spacechar\fi}%
       {\@citeb\@extra@b@citeb}%
     \NAT@date}}
  {\@citea\NAT@nmfmt{\NAT@nm}%
   \NAT@spacechar\NAT@@open\if*#1*\else#1\NAT@spacechar\fi\NAT@hyper@{\NAT@date}}
  {}{}

\makeatother

\AtBeginEnvironment{quote}{\singlespace\vspace{-\topsep}\small}
\AtEndEnvironment{quote}{\vspace{-\topsep}\endsinglespace}
\newif\ifblackandwhite
 \blackandwhitetrue
 \ifblackandwhite
  
\else

\fi

\newcommand{\argmin}{\operatornamewithlimits{arg\,min}}

\newcommand {\what} {\widehat}
\newcommand {\lya} {Ly$\alpha$ }

\newcommand {\HI} {H\hspace{0.5ex}{\scriptsize I}}
\newcommand {\CIV} {C\hspace{0.5ex}{\scriptsize IV}}
\newcommand {\OVI} {{\mbox{O\hspace{0.5ex}{\scriptsize VI }}}}
\newcommand {\MgII} {Mg\hspace{0.5ex}{\scriptsize II }}

\definecolor{block-gray}{gray}{0.98}
\newtcolorbox{blockquote}{colback=block-gray,grow to right by=-1mm,grow to left by=-1mm,boxrule=0pt,boxsep=0pt,breakable}

\title{Trend Filtering -- II. Denoising Astronomical Signals with Varying Degrees of Smoothness}
\author[C. A. Politsch et al.]{Collin A. Politsch,$^{1,2,3}$\thanks{E-mail: capolitsch@cmu.edu}
Jessi Cisewski-Kehe,$^{4}$
Rupert A. C. Croft,$^{3,5,6}$ \newauthor
and Larry Wasserman$^{1,2,3}$
\vspace{.2cm}
\\ 
$^{1}$ Department of Statistics \& Data Science, Carnegie Mellon University, Pittsburgh, PA 15213 \\
$^{2}$ Machine Learning Department, Carnegie Mellon University, Pittsburgh, PA 15213 \\
$^{3}$ McWilliams Center for Cosmology, Carnegie Mellon University, Pittsburgh, PA 15213 \\
$^{4}$ Department of Statistics and Data Science, Yale University, New Haven, CT 06520 \\
$^{5}$ Department of Physics, Carnegie Mellon University, Pittsburgh, PA 15213 \\
$^{6}$ School of Physics, University of Melbourne, VIC 3010, Australia
}

\date{Accepted XXX. Received YYY; in original form 2019 August 20}

\pubyear{2020}

\makeatletter
\def\BState{\State\hskip-\ALG@thistlm}
\makeatother

\begin{document}
\label{firstpage}
\pagerange{\pageref{firstpage}--\pageref{lastpage}}
\maketitle
\title{Denoising Astronomical Signals via Trend Filtering}
\begin{abstract}

Trend filtering---first introduced into the astronomical literature in Paper~I of this series---is a state-of-the-art statistical tool for denoising one-dimensional signals that possess varying degrees of smoothness. In this work, we demonstrate the broad utility of trend filtering to observational astronomy by discussing how it can contribute to a variety of spectroscopic and time-domain studies. The observations we discuss are (1) the Lyman-$\alpha$ forest of quasar spectra; (2) more general spectroscopy of quasars, galaxies, and stars; (3) stellar light curves with planetary transits; (4) eclipsing binary light curves; and (5) supernova light curves. We study the Lyman-$\alpha$ forest in the greatest detail---using trend filtering to map the large-scale structure of the intergalactic medium along quasar-observer lines of sight. The remaining studies share broad themes of: (1) estimating observable parameters of light curves and spectra; and (2) constructing observational spectral/light-curve templates. We also briefly discuss the utility of trend filtering as a tool for one-dimensional data reduction and compression.

\end{abstract}




\section{Introduction}
\label{sec:1}

Many astronomical analyses can be described by the following problem setup. Suppose we collect noisy measurements of an observable quantity (e.g., flux, magnitude, photon counts) according to the data generating process
\begin{equation}
f(t_i) = f_0(t_i) + \epsilon_i,  \hfill i=1,\dots,n,
\end{equation} 
where $t_1,\dots,t_n$ is an arbitrarily-spaced grid of one-dimensional inputs (e.g., times or wavelengths), $\epsilon_i$ is mean zero noise, and $f_0$ is a signal that may exhibit varying degrees of smoothness across the input domain (e.g., a smooth signal with abrupt dips/spikes). Given the observed data sample, we then attempt to estimate (or denoise) the underlying signal $f_0$ from the observations by applying appropriate statistical methods. In \cite{Politsch_2020a}---hereafter referred to as Paper~I---we introduced trend filtering \cite[][]{genlasso,Tibs} into the astronomical literature. When the underlying signal is spatially heterogeneous, i.e. possesses varying degrees of smoothness, trend filtering is superior to popular statistical methods such as Gaussian process regression, smoothing splines, kernels, LOESS, and many others \cite[][]{Nemirovskii_1985, Nemirovskii_1985b,Tibs}. Furthermore, the trend filtering estimate can be computed via a highly efficient and scalable convex optimization algorithm \cite[][]{Ramdas} and only requires data-driven selection of a single scalar hyperparameter. In this paper, we directly demonstrate the broad utility of trend filtering to observational astronomy by using it to carry out a diverse set of spectroscopic and time-domain analyses.

The outline of this paper is as follows. In Section~\ref{sec:lya}, we use trend filtering to study the Lyman-$\alpha$ forest of quasar spectra---a series of absorption features that can be used as a tracer of the matter density distribution along quasar-observer lines of sight. We choose to study this application in depth and then illustrate the breadth of trend filtering's utility through our discussions in Section~\ref{sec:applications}. The applications we discuss in Section~\ref{sec:applications} can be grouped into two broad (and often intertwined) categories: (1) deriving estimates of observable parameters from trend-filtered observations; and (2) using trend filtering to construct spectral/light-curve templates of astronomical objects/events. In Section~\ref{subsec:spectra}, we discuss constructing spectral template libraries for astronomical objects by trend filtering coadded spectroscopic observations. We illustrate our approach on quasar, galaxy, and stellar spectra from the Sloan Digital Sky Survey. Emission-line parameters can also be robustly estimated by fitting radial basis functions (e.g., Gaussians) to trend-filtered estimates near emission lines. In Section~\ref{subsec:exoplanet}, we use relaxed trend filtering to model the detrended, phase-folded light curve of a Kepler stellar system with a transiting exoplanet. We derive estimates and full uncertainty distributions for the transit depth and total transit duration. In Section~\ref{subsec:EB}, we use trend filtering to denoise a detrended, phase-folded Kepler light curve of an eclipsing binary (EB) system. We illustrate that trend filtering provides significant improvements upon the popular \texttt{polyfit} method of \cite{polyfit} that is used to model Kepler EB light curves and derive observable parameters. In Section~\ref{subsec:supernova}, we discuss using trend filtering to construct light-curve templates of supernova (SN) explosions. We illustrate this approach on a SN light curve obtained from the Open Supernova Catalog (\citealt{2017ApJ_OSC}). Furthermore, we derive estimates and full uncertainty distributions for a set of observable parameters---namely, the maximum apparent magnitude, the time of maximum, and the decline rate. Finally, in Section~\ref{subsec:compression}, we briefly discuss a different, non-data-analysis application of trend filtering. Specifically, we discuss the use of trend filtering as a tool for fast and flexible one-dimensional data reduction and compression. The flexibility of the trend filtering estimator, paired with its efficient speed and storage capabilities, make it a potentially powerful tool to include in large-scale (one-dimensional) astronomical data reduction and storage pipelines. 

We utilize the \cite{glmgen} \texttt{glmgen} \textsf{R} implementation of the \cite{Ramdas} trend filtering optimization algorithm in this work. See Paper~I for implementations in other programming languages.

\section{MAIN APPLICATION: QUASAR LYMAN-$\boldsymbol{\alpha}$ FOREST}
\label{sec:lya}

The Lyman-$\alpha$ (Ly$\alpha$) forest is the name given to the absorption features seen in quasar spectra which are caused by neutral hydrogen (\HI) in the intergalactic medium between a quasar and an observer. When emitted from an accretion disk close to the central black hole, the light from the quasar  has a relatively smooth spectrum---a continuum---caused by the summed black-body emission of gas with different temperatures at different disk radii \citep{osterbrock06}. Emission lines are also seen, and their intensities and line ratios supply information on the physical conditions in the line emitting gas. At least twenty broad  emission lines, broadened by high velocities and temperatures can be measured in a single AGN, along with a similar number of narrow lines from colder gas \citep{marziani96}. The emitted spectrum therefore already consists of a superposition of components with varying degrees of smoothness. The Ly$\alpha$ forest arises when this spectrum is further processed with the addition of absorption lines. Light moving towards the observer is redshifted into resonance with the Ly$\alpha$ transition of \HI, and the strength of absorption features is dictated by the densities of intergalactic material along the line of sight \citep{rauch98}. The smoothness of the absorption lines varies depending on the gas pressure, and thermal doppler broadening \citep{gnedin1998,peeples10}. Sharper absorption features, metal lines, are also caused by other intergalactic species, such as \CIV, \OVI and \MgII \citep{hellsten97,pieri2010}. The usefulness of the Ly$\alpha$ forest as a cosmological probe \citep[e.g.,][]{pal2015} stems from its relationship to the matter density field in the Universe, effectively mapping out structure along each quasar-observer line of sight \citep[e.g.,][]{croft97,lee14}. In order to extract this information from noisy spectra and separate it from other components, it is useful to have a method that can deal with the complexities outlined above, i.e. one that can naturally adapt to varying degrees of smoothness without extensive tuning. 

The relative fluctuations in the Ly$\alpha$ forest transmitted flux fraction are of primary interest since they possess a monotonic relationship with the relative distribution of the absorbing \HI. We utilize trend filtering to first denoise the spatially heterogeneous flux signal in an observed \lya forest. Estimates for the fluctuations in transmitted flux due to absorbing \HI\hspace{0.15ex} are then typically produced by coupling the denoised \lya forest with estimates for the quasar continuum and the cosmic mean transmitted flux in the \lya forest. We take an alternative approach: directly estimating the mean flux level---defined as the product of the continuum and cosmic mean transmitted flux---as in \cite{Rupert1}. The mean flux level is a very smooth, spatially homogeneous function within the truncated \lya forest restframe. It is therefore appropriate to use a linear smoother (see Paper~I, Section~2) for this stage of estimation. Specifically, we use local polynomial regression (LOESS; \citealt{Cleveland}; \citealt{Fan1996}; \citealt{Loader}). In this section, we illustrate these methods on a mock quasar \lya forest from \cite{Bautista} and a real quasar \lya forest from the Baryon Oscillation Spectroscopic Survey Data Release 12 (BOSS DR12; \citealt{Alam}) of the Sloan Digital Sky Survey III \cite[SDSS-III;][]{Eisenstein_2011, SDSS-III}.

Historically, \lya forest analyses have typically utilized kernel smoothers \citep[e.g.,][]{Rupert1,Kim_lya}, wavelets \citep[e.g.,][]{Theuns_2000}, or Gaussian processes \cite[e.g.,][]{refId0}. 

\subsection{Notation}
Suppose we observe a quasar located at redshift $z = z_0$. Ignoring systematic effects such as sky contamination and interstellar extinction for the moment, the observational DGP of the \lya forest can be assumed to follow the model
\begin{align}
f(\lambda) &= f_0(\lambda) + \epsilon(\lambda), &\;\;\;\;\;\; \lambda\in\Lambda(z_0), \label{DGP1} \\
&= \overline{F}(\lambda)\cdot C(\lambda)\cdot \big( 1+ \delta_{F}(\lambda)\big) + \epsilon(\lambda), \label{DGP2}
\end{align} 
where $f(\lambda)$ is the observed flux at wavelength $\lambda$, $f_0(\lambda)$ is the flux signal, $\epsilon(\lambda)$ is zero mean white Gaussian noise, $\Lambda(z_0) = (\lambda_{\text{Ly}\beta}, \;\lambda_{\text{Ly}\alpha})\cdot (1+z_0)$ is the redshifted \lya forest, $C(\lambda)$ is the flux of the unabsorbed quasar continuum, $F(\lambda) = f_0(\lambda)/C(\lambda)$ is the transmitted flux fraction, $\overline{F}(\lambda) = \mathbb{E}[F(\lambda)]$ is the mean transmitted flux fraction (over the sky) in the \lya forest at redshift $z = \lambda/\lambda_{\text{Ly}\alpha} - 1$, and 
\begin{equation}
\delta_{F}(\lambda) = F(\lambda)/\overline{F}(\lambda) -1 \label{dF_def}
\end{equation} 
is the fluctuation about the mean \lya transmitted flux at redshift $z = \lambda/\lambda_{\text{Ly}\alpha} - 1$. Here, $\delta_{F}$ is the quantity we are primarily interested in estimating since $\delta_F \propto \delta_{\text{\HI}}^{-1}$ at each fixed redshift, where $\delta_{\text{\HI}}$ is the density of \HI. The estimation of the flux signal $f_0$ is viewed as an ancillary step.

Although, in principle, it is preferable to study the full spectral range $\Lambda(z_0)$ we have found that, in the nonparametric setting, estimating the quasar continuum near the localized \lya and Ly$\beta$ emission peaks at the boundaries of the \lya forest reduces the estimation accuracy in the interior of $\Lambda(z_0)$. Therefore, in this work we limit our analysis to the truncated \lya forest range
\begin{equation}
\overline{\Lambda}(z_0) = (1045 \;\text{\AA}\;, 1195 \;\text{\AA})\cdot (1+z_0). \label{truncatedlya}
\end{equation}
We simplify notation in this work by changing the input space of the functions introduced above by merely altering the input variable. For example, with respect to $\delta_F$, we maintain the notation $\delta_F(\cdot)$ for all inputs $\lambda$, $\nu$, $z$, $\zeta$, while it is understood that a proper change of input spaces has taken place. The various input spaces are defined in Table~\ref{inputs}. 

\begin{table}
\center
\begin{tabular}{c|l|l} 
	\midrule\midrule
	Input & Definition & Range (quasar at $z=z_0$)\\
	\midrule\midrule 
	$\lambda$ & \text{Observed wavelength} & $\overline{\Lambda}(z_0)$\\ 
	$\nu$ & \text{Rest wavelength} & $\overline{\Lambda}_{\text{rest}}(z_0) = \overline{\Lambda}(z_0) / (1+z_0)$\\ 
	$z$ & \text{Redshift} & $\Pi(z_0) = \overline{\Lambda}(z_0) / \lambda_{\text{Ly}\alpha} - 1$ \\
	$\zeta$ & \text{Log-wavelength (scaled)} & $Z(z_0)=10^4\cdot \log_{10}(\overline{\Lambda}(z_0))$ \\ \midrule\midrule
\end{tabular} 
\caption{Various input spaces utilized for the \lya forest analysis. Notation of functions is held constant, e.g. $\delta_F(\cdot)$, and an alteration of the input variable implicitly indicates a change of input spaces. Logarithmic wavelengths are scaled for numerical stability of the trend filtering optimization algorithm.} \label{inputs}
\end{table}

\subsection{Trend filtering the observed flux}
\label{subsec:lyatrendfilter}
We use quadratic trend filtering \cite[][]{genlasso,Tibs} to estimate the flux signal $f_0$ of the observational model (\ref{DGP1}). In both BOSS DR12 and the \cite{Bautista} mock catalog, the quasar spectra are sampled on equally-spaced grids in logarithmic wavelength space with $\Delta\log_{10}(\lambda_i) = 10^{-4}$ dex (in logarithmic angstroms). Furthermore, flux measurement variances are provided by the BOSS pipeline (\citealt{Bolton_2012}), accounting for the statistical uncertainty introduced by photon noise, CCD read noise, and sky-subtraction error. We correct the BOSS spectrum for interstellar extinction with the \cite{Cardelli} extinction law and the \cite{Schlafly} dust map. 

We fit the trend filtering estimator on the equally-spaced logarithmic grid and tune the complexity by minimizing Stein's unbiased risk estimate (SURE) of the fixed-input mean-squared error (see Paper~I). More precisely, we fit the trend filtering estimator in the input space $Z(z_0)=10^4\cdot \log_{10}(\overline{\Lambda}(z_0))$, as defined in Table \ref{inputs}, where we add the scaling to unit spacing for numerical stability of the trend filtering convex optimization.

\subsection{Nonparametric continuum estimation}
\label{subsec:dF}

We utilize a modified \cite{Rupert1} approach to propagate the trend filtering estimate for the flux signal $f_0$ from Section~\ref{subsec:lyatrendfilter} into an estimate for the fluctuation field $\delta_F$ along a line of sight to an observed quasar. Namely, given the trend filtering estimate $\what{f}_0$, we directly estimate the smooth mean flux level defined as the product $m = \overline{F}\cdot C$ and then define the $\delta_F$ estimates via the transformation $\what{\delta}_F := \what{f}_0/\what{m} - 1$. We carry out the estimation of $m$ via a wide-kernel LOESS smooth of the trend filtering estimate, with the specific bandwidth of the kernel selected by optimizing over a large sample of mock spectra (detailed in Section~\ref{bandwidth}). We find that regressing on the fitted values of the trend filtering estimate---instead of the observational DGP (\ref{DGP1})---significantly improves the accuracy and robustness of the $\delta_F$ estimates. We carry out the LOESS estimation in the \lya restframe $\overline{\Lambda}_{\text{rest}}(z_0)$ (see Table \ref{inputs}) in order to remove the effect of redshifting on the smoothness of $m$. The LOESS estimation of $m$ is a fully nonparametric procedure and provides a reduction in bias over popular parametric approaches such as low-order power laws and principal components analyses (PCA). The sole assumption of our LOESS approach is that, in the restframe, the mean flux level $m$ always has a fixed degree of smoothness, defined by an optimal fixed kernel bandwidth. There is of course a bias-variance tradeoff here; namely, the decreased bias comes with a modest increase in variance compared to a parametric power law or low-dimensional PCA model. Our decision in favor of the LOESS approach directly reflects our stance that low bias is preferable to low variance in this context since statistical uncertainty due to estimator variability is tracked by our uncertainty quantification (Section~\ref{subsec:lya_uc}), while uncertainty due to modeling bias is not easily quantifiable. Therefore, we can be more confident that significant fluctuations in the estimated $\delta_F$ field are in fact real, and not due to statistical bias in the quasar continuum estimate.

To be explicit, the LOESS estimator for $m$ is a regression on the data set
\begin{equation}
\{(\nu_i,\;\what{f}_0(\nu_i; \what{\gamma})\}_{i=1}^{n},\hfill \nu_i\in \overline{\Lambda}_{\text{rest}}(z_0), \label{TFpairs}
\end{equation}
which can be viewed as arising from the DGP
\begin{equation}
\what{f}_0(\nu_i; \what{\gamma}) = m(\nu_i) + \rho_i, 
\end{equation} 
where $\what{f}_0$ is the trend filtering estimate fixed at the minimum SURE hyperparameter $\what{\gamma}$, $e_i = \what{f}_0(\nu_i; \what{\gamma}) - f_0(\nu_i)$ are the errors of the trend filtering estimate, and $\rho_i = m(\nu_i)\cdot \delta_{F}(\nu_i) + e_i$ are autocorrelated fluctuations about zero. The LOESS estimator is the natural extension of kernel regression (\citealt{Nadaraya}; \citealt{Watson}) to higher-order local polynomials. Given a kernel function $K(\cdot)$ with bandwidth $h>0$, for each $i=1,\dots,n$, the LOESS estimator is obtained by minimizing\begin{equation}
\sum_{j=1}^{n}{\Big(\what{f}_0(\nu_j; \what{\gamma}) - \phi_{\nu_i}(\nu_{j};\beta_0,\dots,\beta_d)\Big)^2}K\Bigg(\frac{|\nu_j - \nu_i|}{h}\Bigg),
\end{equation}
and letting $\widehat{m}(\nu_i) = \what{\beta}_0$, where $\phi_{\nu_i}(\cdot\;;\beta_0,\dots,\beta_d)$ is a $d$th order polynomial centered at $\nu_i$. Specifically, we utilize the local linear regression estimator (LLR; $d=1$) and the Epanechnikov kernel (\citealt{Epanechnikov})
\begin{equation}
K(t) = \frac{3}{4}(1-t^2)\mathbbm{1}\{|t|<1\}.
\end{equation}
The LLR estimator is described in full detail by Algorithm \ref{Alg4}. Given the trend filtering estimate $\what{f}_0$ and the LLR estimate $\what{m}$, the $\delta_F$ estimates are then defined as 
\begin{equation}
\what{\delta}_{F}(z_i; \what{\gamma}) = \frac{\what{f}_0(z_i; \what{\gamma})}{\what{m}(z_{i})} - 1, \hfill z_1,\dots,z_n\in\Pi(z_0), \label{dF_estimates}
\end{equation}
where we deliberately express $\what{m}$ as ``hyperparameter-less'' since $\gamma$ has already been fixed at the minimum SURE value $\what{\gamma}$ and we provide the optimal LOESS bandwidth value $h_0 = 74$ \AA---optimized over the large mock sample. Here, we have also done a change of variables to redshift space---our desired input domain for studying the \HI\hspace{0.15ex} density fluctuations in the IGM.

\begin{algorithm}
\caption{\small LOESS (local linear) estimator for mean flux level}
\begin{algorithmic}[1]
\REQUIRE { Training Data $\{(\nu_i,\what{f}_0(\nu_i; \what{\gamma}))\}_{i=1}^{n}$, $\;$Bandwidth $h_0 = 74$ \AA} 
\FORALL{$i$}
	\STATE Let $\what{\beta}_0$, $\what{\beta}_1$ minimize the locally weighted sum of squares
	\begin{equation*}
	\sum_{j=1}^{n}{\Big(\what{f}_0(\nu_j; \what{\gamma}) - \beta_0 - \beta_1(\nu_j-\nu_i)\Big)^2}\cdot K\Bigg(\frac{|\nu_j - \nu_i|}{h_0}\Bigg). 
	\end{equation*}
	\STATE Let $\what{m}(\nu_i)  = \what{\beta}_0(\nu_i)$.
\ENDFOR \\
\hspace{-3.75ex}{\bf Output:} $\{\what{m}(\nu_i)\}_{i=1}^{n}$
\end{algorithmic} \label{Alg4}
\end{algorithm}

Although $h_0$ is chosen to directly optimize $\what{\delta}_F$ accuracy, an estimate for the quasar continuum $C(\cdot)$ arises intrinsically: 
\begin{equation}
\what{C}(\nu_i) = \overline{F}(\nu_i)^{-1}\cdot \what{m}(\nu_i),\hfill \nu_i\in\overline{\Lambda}_{\text{rest}}(z_0),
\end{equation}
where precise estimates of $\overline{F}$ follow from a rich literature (e.g. \citealt{Bernardi_2003}; \citealt{McDonald3}; \citealt{Faucher_2008}; \citealt{DallAglio_2009}; \citealt{Paris}; \citealt{Becker_2013}). The $\delta_F$ estimates could then be equivalently restated as
\begin{equation}
\what{\delta}_F(z_i; \what{\gamma}) = \frac{\what{F}(z_i; \what{\gamma})}{\overline{F}(z_i)} - 1, \hfill z_1,\dots,z_n\in\Pi(z_0),
\end{equation}
where
\begin{equation}
\what{F}(z_i; \what{\gamma}) = \what{f}_0(z_i; \what{\gamma})/\what{C}(z_i). \label{transmittedflux}
\end{equation}

\subsection{Calibrating continuum smoothness}
\label{bandwidth}
\begin{figure}
    \begin{center}
    \includegraphics[width = 0.475\textwidth]{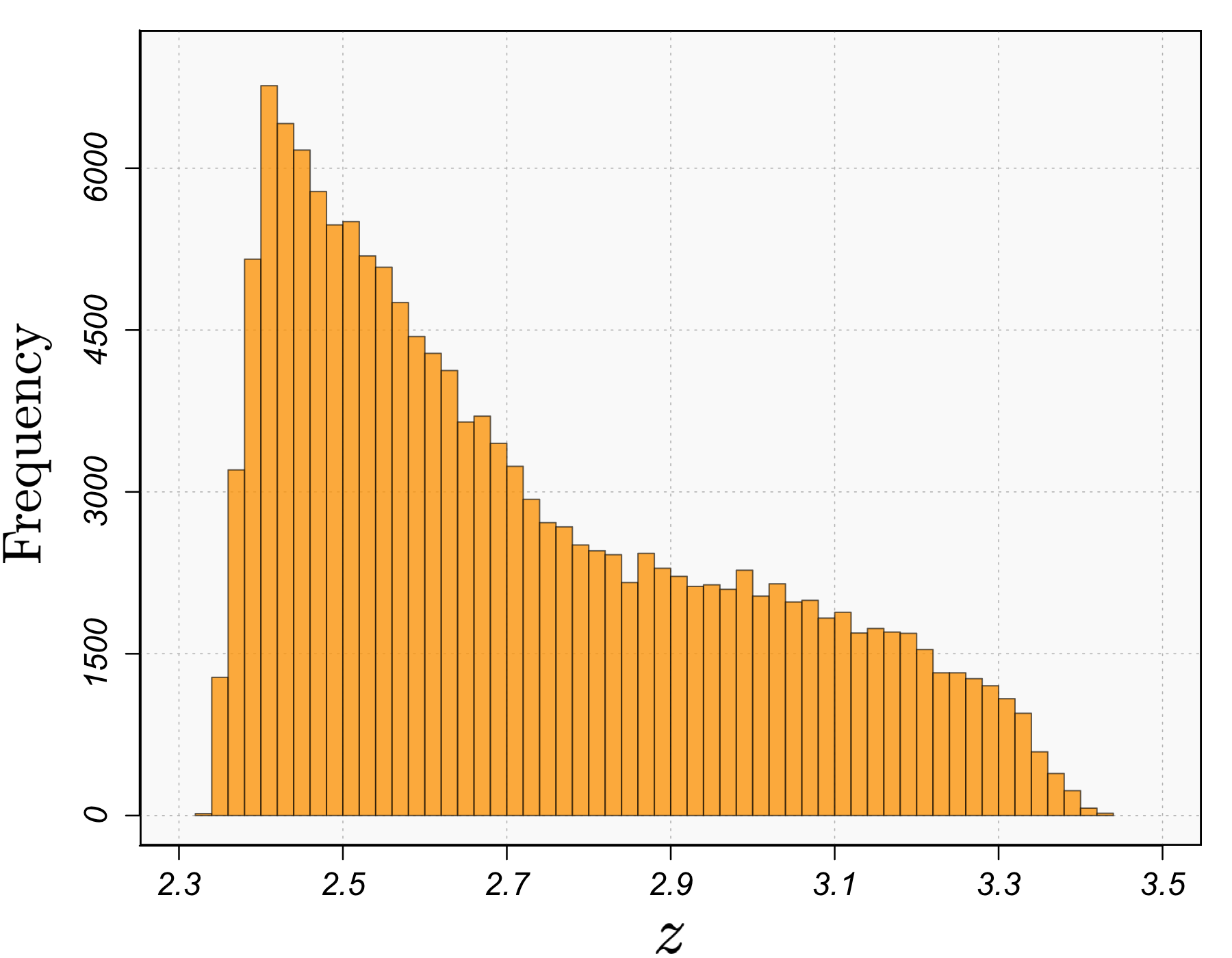} \\
    \includegraphics[width=.475\textwidth]{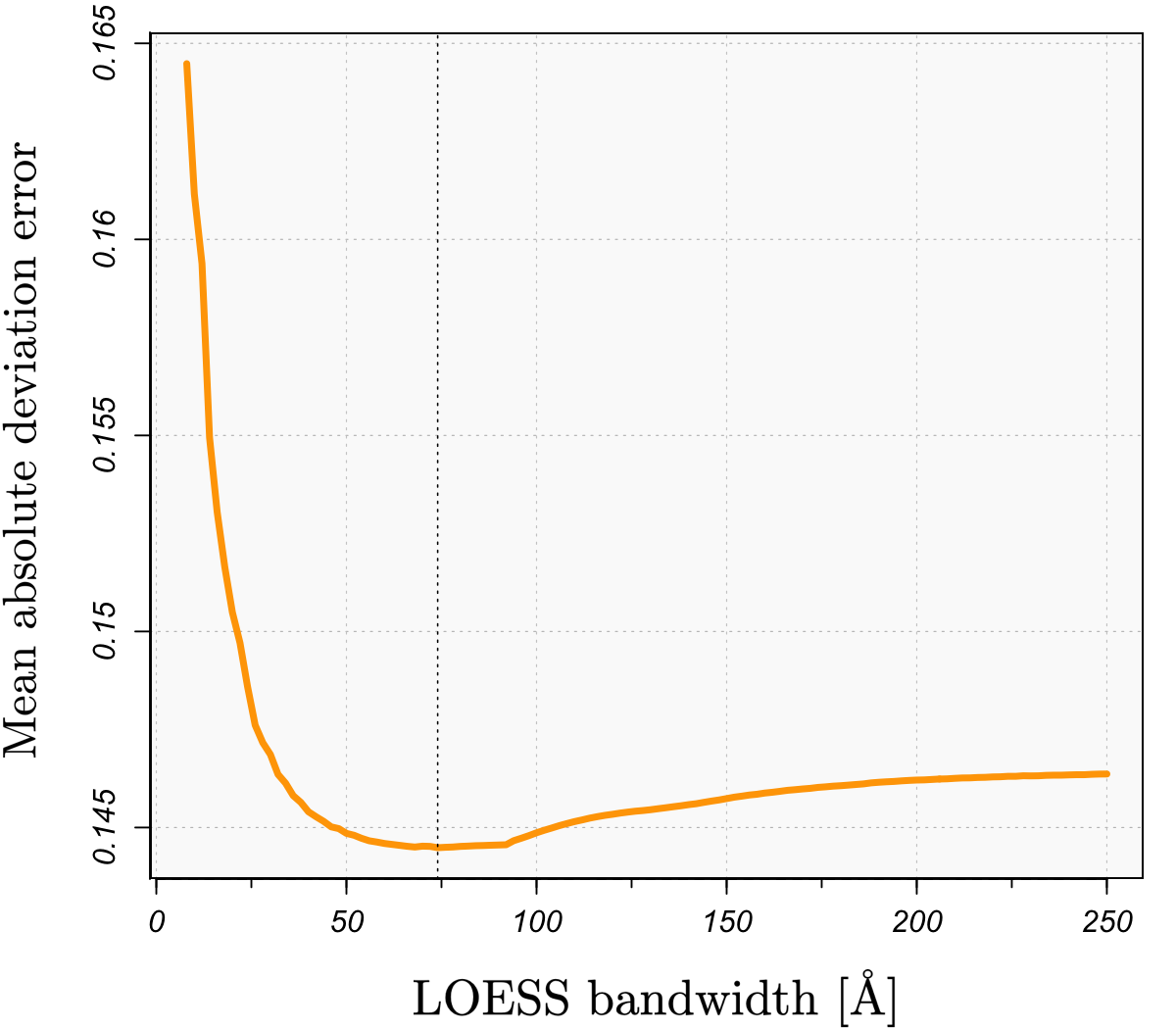}
    \vspace{-0.25cm}
    
    \caption{{\bf Top}: Distribution of mock quasar redshifts (data reduction detailed in the appendix). We utilize this sample of 124,709 quasars to calibrate the optimal nonparametric continuum smoothness. {\bf Bottom}: Mean absolute deviation error curve for selecting the optimal kernel bandwidth for the LOESS (local linear) estimator of the mean flux level, averaged over the 124,709 spectra in the mock sample. The optimal choice of bandwidth is $h_0 = 74$ \AA.} \label{h_risk}
    \end{center}
\end{figure}

We utilize a sample of 124,709 mock quasar spectra from the \cite{Bautista} catalog to optimize the bandwidth of the LOESS estimator for the mean flux level that intrinsically removes the effect of the quasar continuum. Our mock data reduction is detailed in the appendix and the redshift distribution of the quasars is shown in the top panel of Figure~\ref{h_risk}.

For each mock quasar \lya forest with DGP $Q_j$, $j=1$, $\dots$, 124,709, we first compute the trend filtering hyperparameter value that minimizes the \emph{true} fixed-input mean-squared prediction error
\begin{equation}
\gamma_0^j = \argmin_{\gamma>0} \frac{1}{n}\sum_{i=1}^{n}\mathbb{E}_{Q_j}\Big[\big(f(\zeta_i) - \widehat{f}_0(\zeta_i;\gamma)\big)^2\;\Big|\;\zeta_1,\dots,\zeta_n\Big]. \label{gamma0}
\end{equation}
Then, given the trend filtering restframe fitted values $\{(\nu_i,\;\what{f}_0(\nu_i; \gamma_0^j)\}_{i=1}^{n}$, we fit a LOESS estimator with bandwidth $h$ and define the error (as a function of $h$) of the resulting $\delta_F$ estimator as the fixed-input mean absolute deviation (MAD) error
\begin{equation}
R_j(h) = \frac{1}{n}\sum_{i}\mathbb{E}_{Q_j}\Big[\big|\delta_{F}(\nu_i) - \widehat{\delta}_F(\nu_i; \gamma_0^j,h)\big|\;\Big|\;\nu_1,\dots,\nu_n\Big], \label{risk}
\end{equation}
where $\mathbb{E}_{Q_j}$ denotes the mathematical expectation over the randomness arising from the observational DGP $Q_j$. Because we can repeatedly sample from each mock quasar DGP, the expectations in (\ref{gamma0}) and (\ref{risk}) can be computed to an arbitrary precision. Here, we utilize 300 realizations of each DGP to approximate the mathematical expectations.

We then define the optimal choice of $h$ as the minimizer of the summed error over the full sample of $m = $124,709 mock quasar spectra:
\begin{equation}
h_0 = \argmin_{h>0} \sum_{j=1}^{m} R_j(h).
\end{equation}
The aggregate error curve is shown in the bottom panel of Figure~\ref{h_risk}, yielding an optimal value of $h_0 = 74$ \AA. We find that defining $R(h)$ as the conditional MAD error---instead of the conditional MSE---provides an essential boost in robustness that keeps the error from being dominated by a very small proportion of worst-case estimates. More complex choices of bandwidth, e.g. variable bandwidths that depend on the S/N ratio of the trend filtering estimator, the restframe pixel spacing, and/or the redshift of the quasar, do not significantly improve upon the $h_0 = 74$ \AA$\;$restframe fixed bandwidth.

\subsection{Uncertainty quantification}
\label{subsec:lya_uc}

Given the assumed noise model $\epsilon_i \sim N(0, \sigma_i^2)$, $i=1,\dots,n$ provided by the BOSS pipeline (\citealt{Bolton_2012}), we can construct a pointwise variability band for $\what{\delta}_F$ via an augmentation of the parametric bootstrap outlined in Algorithm~2 of Paper~I. Specifically, given the parametric bootstrap ensemble of trend filtering estimates provided by the parametric bootstrap algorithm, we fit the mean flux level of each with the LOESS estimator detailed in Algorithm \ref{Alg4}, and then define the $\what{\delta}_F$ bootstrap ensemble
\begin{equation}
\what{\delta}_{F,b}^{*}(z_i) = \frac{\what{f}_b^{*}(z_i)}{\what{m}_b^{*}(z_i)} - 1, \hfill i=1,\dots,n,\;\;b=1,\dots,B,
\end{equation}
where, for each $b=1,\dots,B$, $\what{m}_b^{*}$ is the LOESS estimate fit to the data set $\big\{(\nu_i, \what{f}_b^{*}(\nu_i))\big\}_{i=1}^{n}$. Note that refitting the LOESS estimator on each bootstrap realization allows us to track the extra variability introduced into the $\delta_F$ estimates by the uncertainty in the continuum estimate. A $(1-\alpha)\cdot100\%$ quantile-based pointwise variability band for $\what{\delta}_F$ is then given by
\begin{equation}
V_{1-\alpha}(z_i) = \Big(\widehat{\delta}_{F,\alpha/2}^{*}(z_i), \;\widehat{\delta}_{F,1-\alpha/2}^{*}(z_i)\Big), \hfill i = 1,\dots,n. \label{var2}
\end{equation}
where
\begin{equation}
\widehat{\delta}_{F,\beta}^{*}(z_i) = \inf_{g}\Bigg\{g : \frac{1}{B} \sum_{b=1}^{B} \mathbbm{1}\big\{\widehat{\delta}_{F,b}^{*}(z_i) \leq g\big\} \geq \beta \Bigg\},
\end{equation}
for any $\beta\in(0,1)$.

\subsection{Results}

\begin{figure*}
\centering
\includegraphics[width = \textwidth]{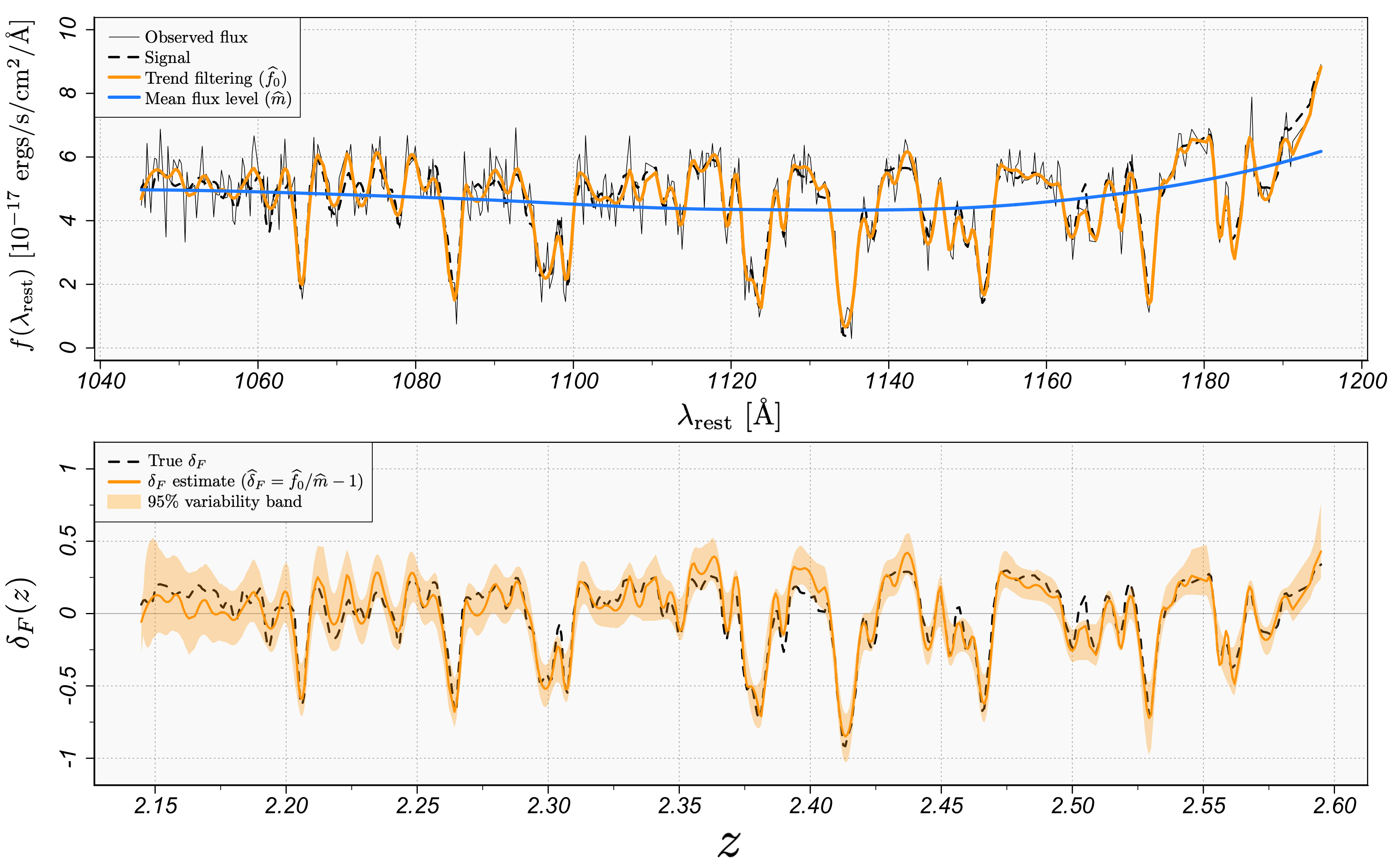} \\
\vspace{-0.75cm}

\includegraphics[width = \textwidth]{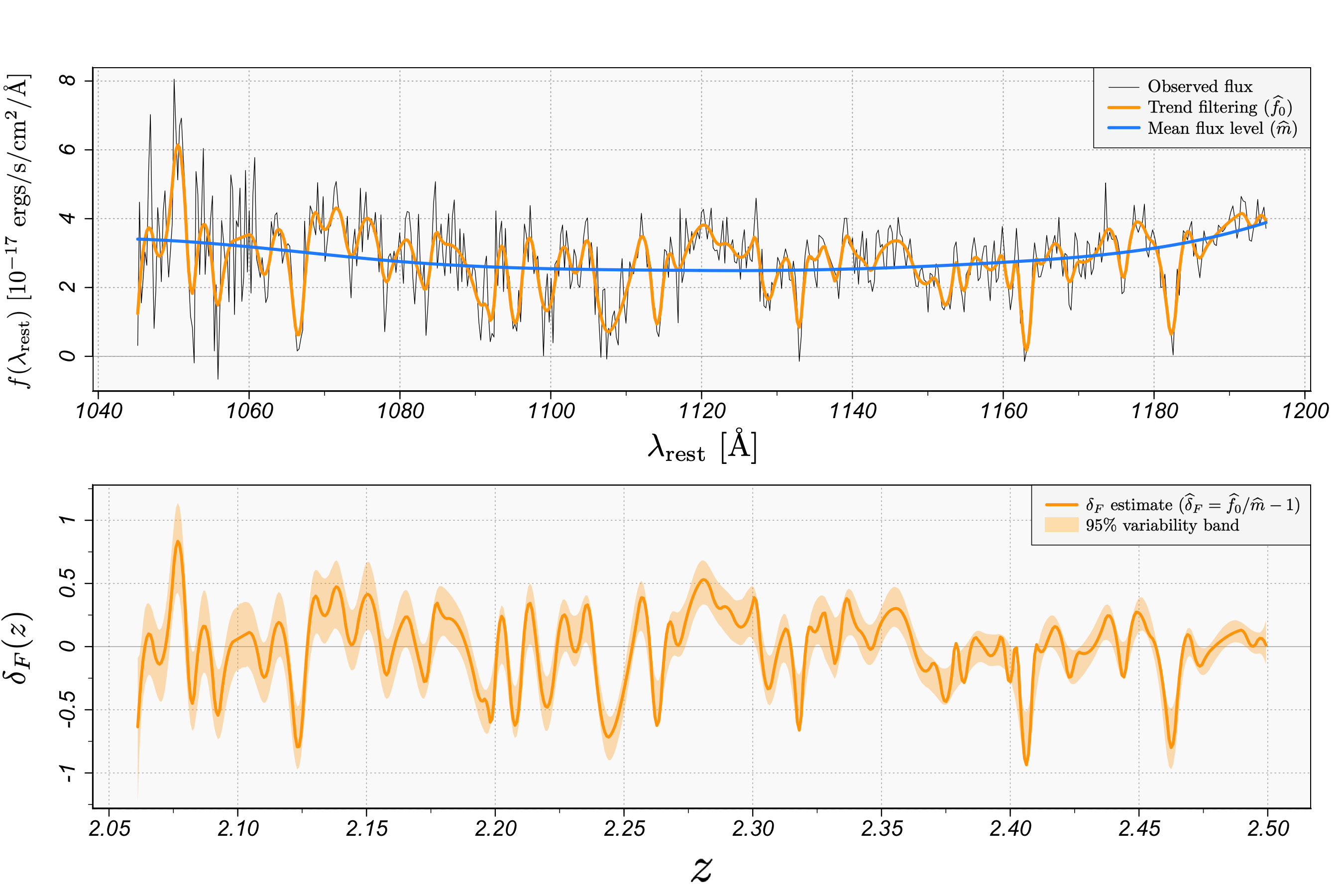}
\vspace{-0.55cm}

\caption{Results of \lya forest analysis. {\bf Top panel}: \lya forest of a mock quasar spectrum (\citealt{Bautista}) in the restframe, with the quadratic trend filtering estimate shown in orange and the LOESS (local linear) estimate for the mean flux level shown in blue. {\bf Second panel}: The redshift-space fluctuations in the \lya transmitted flux fraction, with our estimate superposed. The fluctuations inversely trace the relative under- and over-densities of \HI\hspace{0.15ex} in the intervening intergalactic medium between Earth and the quasar. {\bf Third and Fourth panels}: Analogous plots for a real quasar \lya forest from the twelfth data release of the Baryon Oscillation Spectroscopic Survey (\citealt{Alam}; Plate = 6487, MJD = 56362, Fiber = 647). The quasar is located at (RA, Dec, $z$) $\approx$ (196.680$^{\circ}$, 31.078$^{\circ}$, 2.560). }
\label{sample_lya}
\end{figure*}

A mock quasar spectrum from \cite{Bautista} and a real quasar spectrum from the Baryon Oscillation Spectroscopic Survey (BOSS; \citealt{Alam}) are displayed in Figure~\ref{sample_lya}, with the results of our \lya forest analysis overlaid. Starting with the top panel, the quadratic trend filtering estimate is fit on the equally-spaced observations in logarithmic wavelength space and then transformed to the restframe wavelength space, where the LOESS estimate for the mean flux level---the product of the continuum and cosmic mean \lya absorption---is then fit to the trend filtering fitted values. The $\delta_F$ estimates are then computed according to (\ref{dF_estimates}) and displayed in redshift space in the second panel, where they closely track the true $\delta_F$ defined by (\ref{dF_def}). Furthermore, the 95\% bootstrap variability band defined in (\ref{var2}) can be seen to almost fully cover the true $\delta_F$ signal. The estimated $\delta_F$ can be interpreted as an inversely proportional proxy for the deviations from the mean \HI\hspace{0.15ex} density in the intervening intergalactic medium at each redshift---with negative values of $\delta_F$ corresponding to (epoch-relative) over-densities of \HI\hspace{0.15ex} and positive values corresponding to under-densities. 

The third and fourth panels are the analogous plots for a BOSS quasar spectrum \lya forest (Data Release 12, Plate = 6487, MJD = 56362, Fiber = 647). The quasar is located in the northern galactic cap at (RA, Dec, $z$) $\approx$ (196.680$^{\circ}$, 31.078$^{\circ}$, 2.560).

\section{Further Applications}
\label{sec:applications}

The analysis of signals possessing varying degrees of smoothness permeates many areas of astronomy. In this section, we discuss a variety of further applications for which trend filtering may find use. For the sake of brevity, we discuss these applications in less detail than the \lya forest analysis in Section~\ref{sec:lya}. Naturally, we expect trend filtering may find many uses in astronomy beyond those we explicitly discuss.

\subsection{Spectral template generation and estimation of emission-line parameters}
\label{subsec:spectra}

Automated spectral classification and redshift estimation pipelines require rich template libraries that span the full space of physical objects in the targeted set in order to achieve high statistical accuracy. In this section, we discuss using trend filtering to construct spectral templates from observational spectra. We describe the procedure here for generating a single spectral template from a well-sampled observed spectrum. Suppose we observe coadded flux measurements of a targeted source at wavelengths $\lambda_1,\dots,\lambda_n\in \Lambda$ according to the data generating process
\begin{equation}
f(\lambda_i) = f_0(\lambda_i) + \epsilon_i,  \hfill i=1,\dots,n,
\label{temp_DGP}
\end{equation}
where the observations have been corrected for systematic effects (e.g., sky subtraction, interstellar extinction, etc.) and the $\epsilon_i$ are mean-zero errors that arise from instrumental uncertainty and systematic miscalibrations. After removal of potentially problematic pixels (e.g., near bright sky lines), let $\what{f}_0(\lambda),\;\lambda\in\Lambda$ denote the continuous trend filtering estimate fit to the observations. Given a confident object classification and redshift estimate $z_0$ (e.g., determined by visual inspection), we then define the restframe spectral template
\begin{equation}
b(\lambda_{\text{rest}}) = \what{f}_0(\lambda / (z_0 + 1)), \hfill \lambda_{\text{rest}}\in\Lambda / (z_0 + 1),
\end{equation}
and store it in the respective object template library.

\begin{figure*}
\centering
\includegraphics[width = \textwidth]{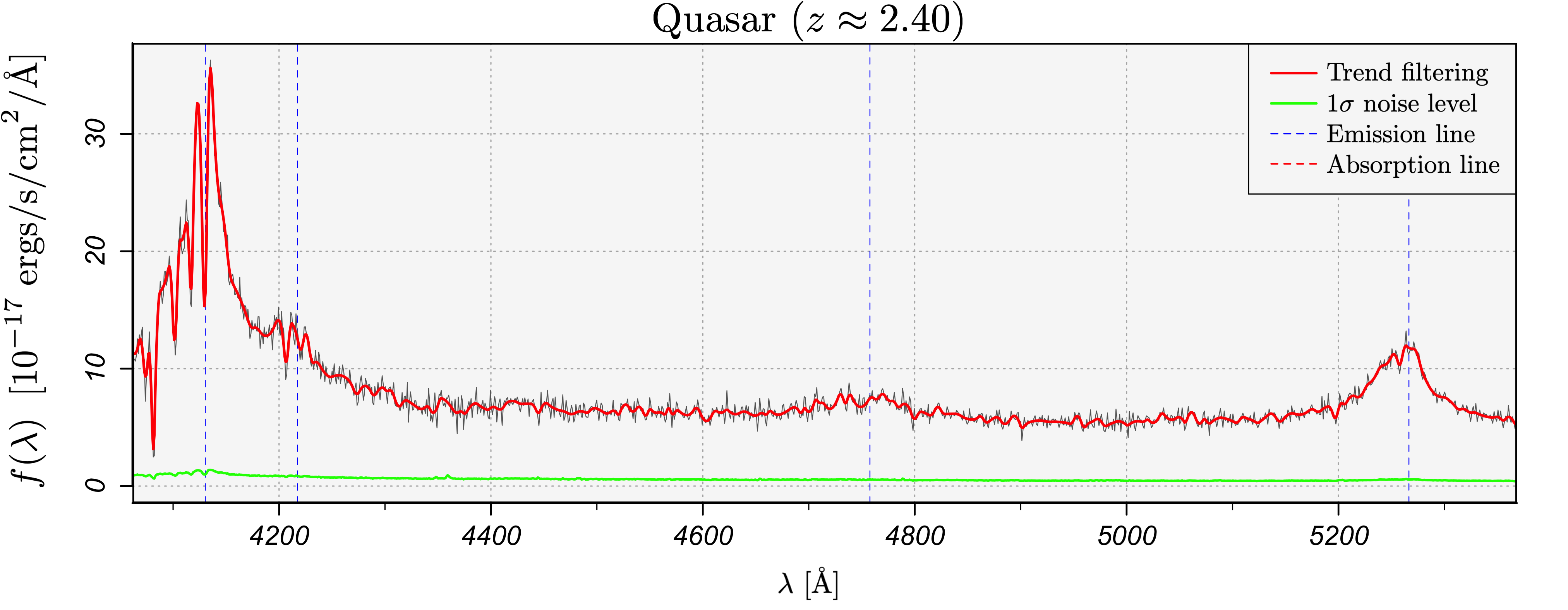} \\
\vspace{0.15cm}

\includegraphics[width = \textwidth]{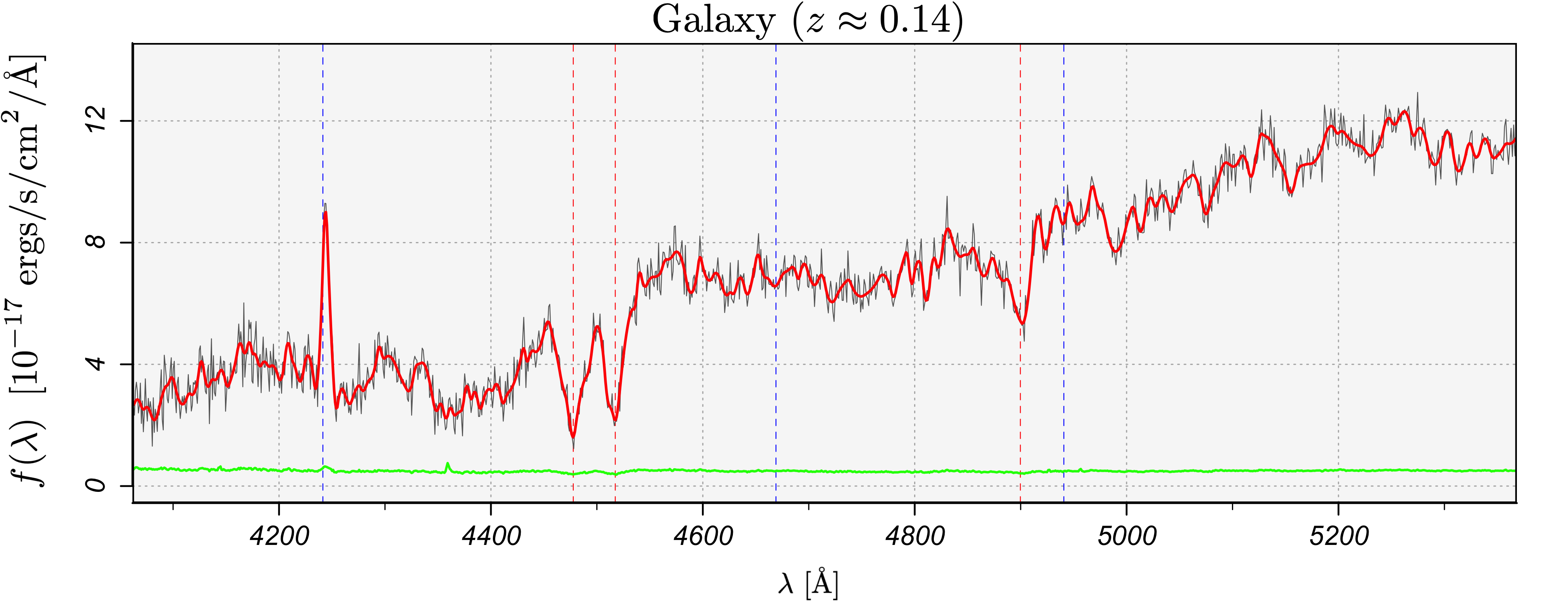} \\
\vspace{0.15cm}

\includegraphics[width = \textwidth]{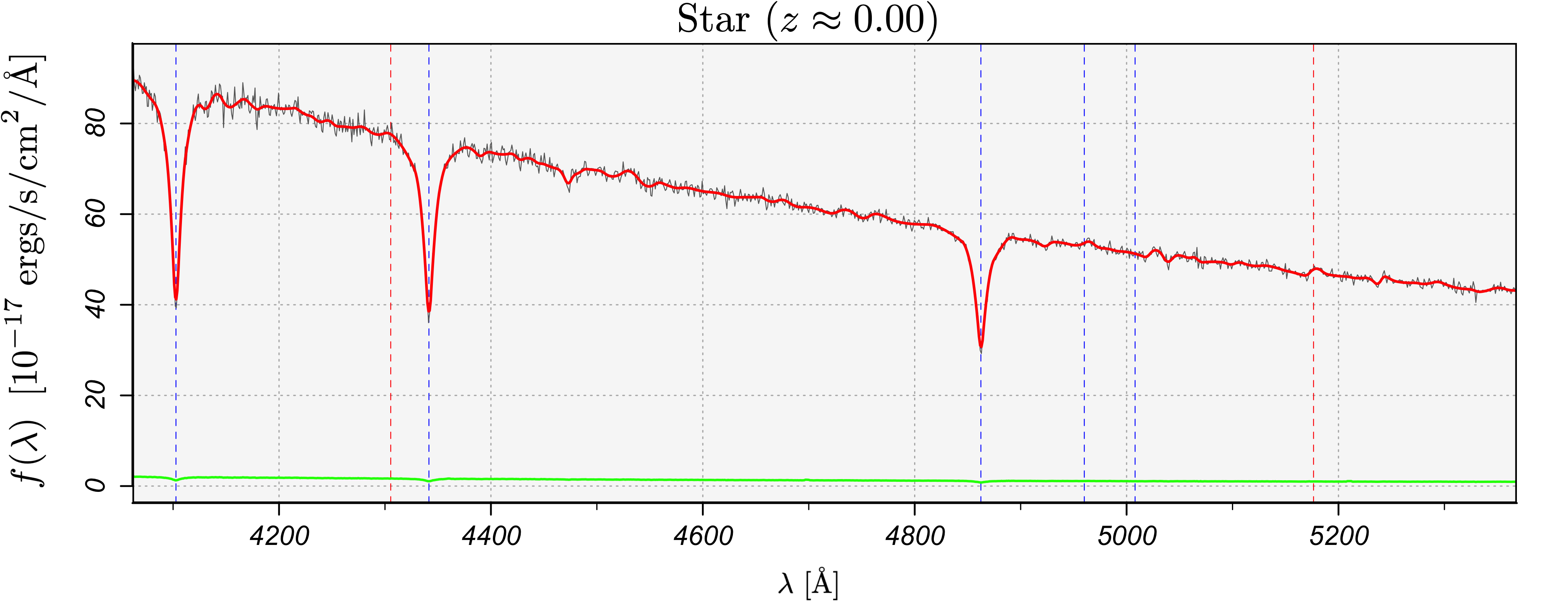} 
\vspace{-0.5cm}

\caption{Optical coadded spectra collected by the Baryon Oscillation Spectroscopic Survey (\citealt{Alam}) of the Sloan Digital Sky Survey III (\citealt{Eisenstein_2011}; \citealt{SDSS-III}). From top to bottom, a quasar (DR12, Plate = 7140, MJD = 56569, Fiber = 58) located at (RA, Dec, $z$) $\approx$ (349.737$^\circ$, 33.414$^\circ$, $2.399$), a galaxy (DR12, Plate = 7140, MJD = 56569, Fiber = 68) located at (RA, Dec, $z$) $\approx$ (349.374$^\circ$, 33.617$^\circ$, $0.138$), and a star (DR12, Plate = 4055, MJD = 55359, Fiber = 84) located at (RA, Dec, $z$) $\approx$ (236.834$^\circ$, 0.680$^\circ$, $0.000$). We fit a quadratic trend filtering estimate to each spectrum in the logarithmic wavelength space in which the observations are equally spaced, and optimize the hyperparameter by minimizing Stein's unbiased risk estimate. Given confidently determined redshifts (e.g., determined by visual inspection), the trend filtering estimate for each object can be scaled to the restframe and stored as a spectral template. Furthermore, emission-line parameter estimates for a spectrum can be obtained by fitting Gaussian radial basis functions to the emission lines of the trend filtering estimate.}
\label{sample_templates}
\end{figure*}

In Figure~\ref{sample_templates}, we show three optical coadded spectra from the twelfth data release of the Baryon Oscillation Spectroscopic Survey (BOSS DR12; \citealt{Alam}) of the Sloan Digital Sky Survey III (SDSS-III; \citealt{Eisenstein_2011}; \citealt{SDSS-III}). The figure is zoomed to a subinterval of the optical range for visual clarity and the spectra are easily identifiable as a quasar, a galaxy, and a star, respectively. We fit an error-weighted quadratic trend filtering estimate to each spectrum in the logarithmic-angstrom wavelength space in which the BOSS spectra are gridded, and tune the hyperparameter to minimize Stein's unbiased risk estimate (see Section~3.5 of Paper~I). The trend filtering estimates give good results, adequately adapting to even the strongest emission and absorption features in each spectrum. 

As in the BOSS spectroscopic pipeline \cite[][]{Bolton_2012}, after a spectral classification and redshift measurement have been precisely determined, emission-line parameter estimates can then be obtained by fitting Gaussian radial basis functions to the emission lines of the spectrum ``best fit''---nonlinearly optimizing the amplitudes, centroids, and widths. We propose that the trend filtering estimate of a spectrum be used as the ``best fit'' for this type of procedure, e.g. instead of the low-dimensional principal components models currently used by the BOSS pipeline. The relative magnitudes of the emission-line parameter estimates can then be used to determine object subclassifications.

\subsection{Exoplanet transit modeling}
\label{subsec:exoplanet}

In this section we discuss the use of trend filtering for modeling the photometric time series of an extrasolar planet transiting its host star. Given a phase-folded stellar light curve, corrected for stellar variability and spacecraft motion systematics, trend filtering can be used to automatically produce fully nonparametric estimates and uncertainties for the transit depth and total transit duration. 

\begin{figure*}
\centering
\includegraphics[width = \textwidth]{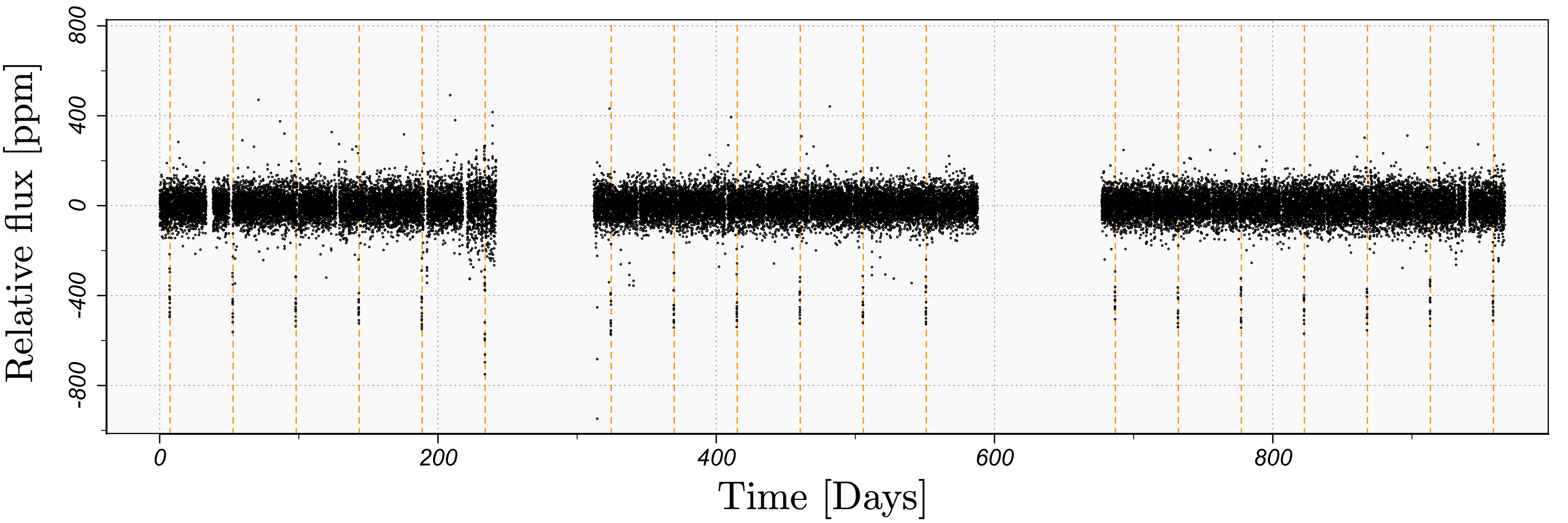} \\
\vspace{0.25cm}

\includegraphics[width = 0.69\textwidth]{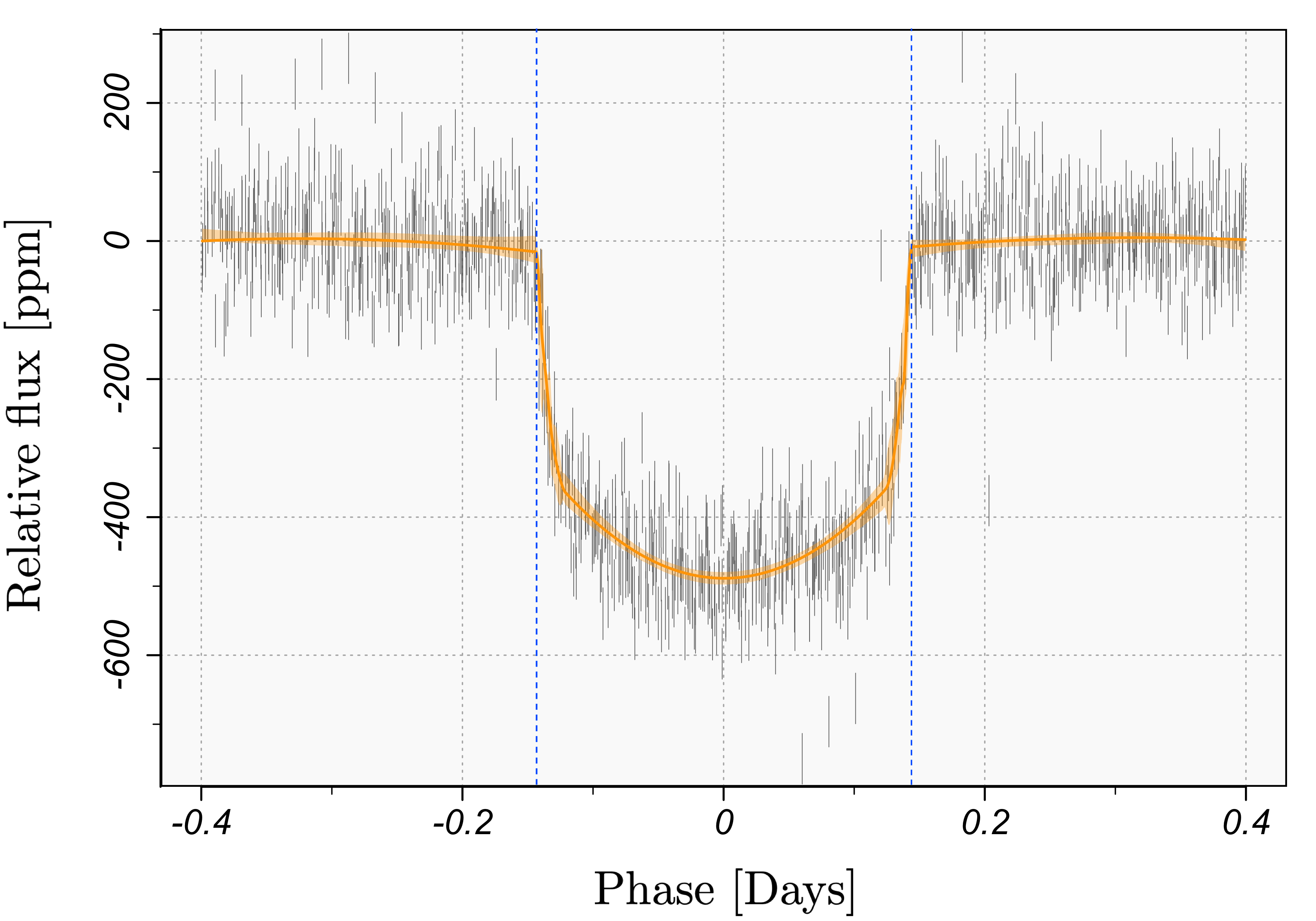} \\
\vspace{0.15cm}

\hspace{0.02\textwidth}\includegraphics[width = 0.45\textwidth]{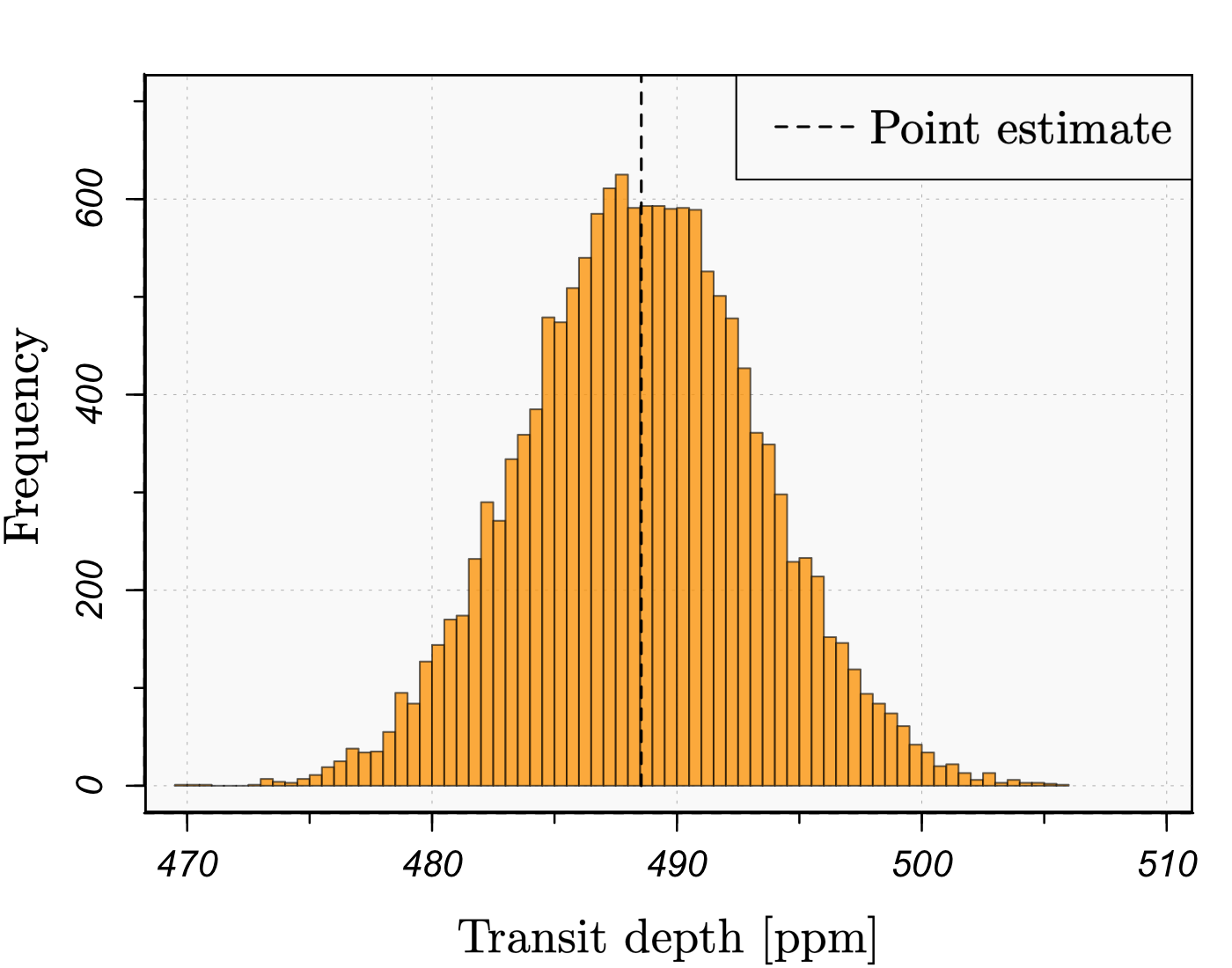}
\includegraphics[width = 0.45\textwidth]{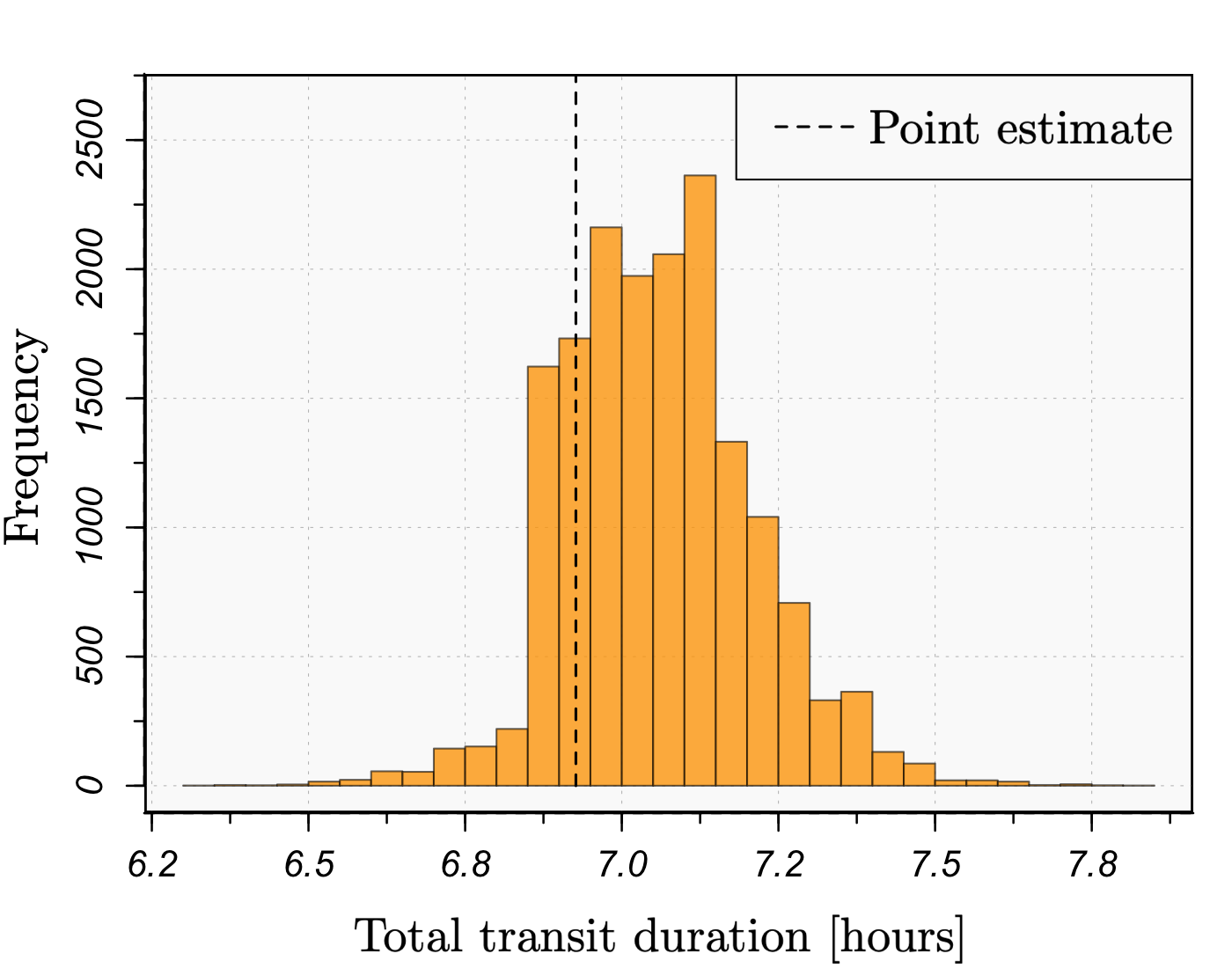}
\caption{Kepler-10c transit light curve analysis. {\bf Top}: Long-cadence (30-min.), quarter-stitched, median detrended, relative flux light curve (\texttt{LC\_DETREND}) of the confirmed exoplanet host Kepler-10 (KOI-072, KIC 11904151; \citealt{kepler10b}), processed by the Kepler pipeline (\citealt{Jenkins_2010}) and obtained from the NASA Exoplanet Archive (\citealt{Akeson_2013}). Vertical lines indicate the observed transit events of the system's second confirmed planet Kepler-10c (KOI-072 c, KIC 11904151 c; \citealt{Fressin_2011}). {\bf Middle}: Phase-folded transit light curve for Kepler-10c ($\sim$45.29 day orbital period) with $1\sigma$ error bars. The error-weighted relaxed trend filtering estimate, optimized by sequential $K$-fold cross validation, is superposed with 95\% variability bands. The estimated inception and termination of the transit event are indicated by the vertical dashed lines. The estimated transit depth and total transit duration are $\what{\delta} = 488.292$ ppm and $\what{T} = 6.927$ hours, respectively. {\bf Bottom}: Bootstrap sampling distributions of the transit depth and transit duration estimates.}
\label{sample_exoplanet}
\end{figure*}

We demonstrate our approach on long-cadence photometric observations of Kepler-10 (KOI-72, KIC 11904151; \citealt{kepler}). Kepler-10 is confirmed to host at least two exoplanets: Kepler-10b (KOI-72 b, KIC 11904151 b; \citealt{kepler10b}) and Kepler-10c (KOI-72 c, KIC 11904151 c; \citealt{Fressin_2011}). Each planet was first detected via the transit method---a measurable periodic dimming in the photometry caused by the planet crossing in front of the host star. Here, we use trend filtering to estimate a nonparametric transit model for Kepler-10c and derive depth and duration measurements. The results of our analysis are displayed in Figure~\ref{sample_exoplanet}. The top panel displays a sample of the Kepler-10 long-cadence (30-min. increment), quarter-stitched, median detrended, relative flux light curve processed by the Kepler pipeline \cite[][]{Jenkins_2010}, which we accessed from the NASA Exoplanet Archive \cite[][]{Akeson_2013}. The observed transit events of Kepler-10c are indicated by the vertical dashed lines. The middle panel displays the light curve (with $1\sigma$ error bars) after phase-folding with respect to the $\sim$45.29 day orbital period of Kepler-10c and zooming in on the transit event. We fit a relaxed quadratic trend filtering estimate (see Section~3.7 of Paper~I) to the phased data, weighted by the measurement variances provided by the Kepler pipeline and tuned by sequential $K$-fold cross validation. The optimal relaxation hyperparameter is $\what{\phi} = 0.14$, indicating that relaxation significantly improves the traditional trend filtering estimate in this setting. In particular, we find that relaxation allows the estimate to faithfully capture the sharp transitions corresponding to the beginning of the ingress phase and the end of the egress phase. The relaxed trend filtering estimate is overlaid on the phase-folded light curve, along with 95\% variability bands. Estimates for the transit depth and total transit duration follow immediately from the relaxed trend filtering estimate, as detailed below. The estimated inception and termination of the transit event are indicated by the vertical dashed lines in the middle panel. The nonparametric bootstrap sampling distributions (see Algorithm~1 of Paper~I) of the transit depth and total transit duration measurements are displayed in the bottom panel. Our point estimates for the transit depth and total transit duration for Kepler-10c are $\what{\delta} = 488.292$ ppm and $\what{T} = 6.927$ hours, respectively.

The knot-selection property of trend filtering is particularly appealing in this setting because it provides interpretation as to where the transit event begins and ends. Specifically, we define our estimate of the inception of the transit event $\what{T}_0$ as the leftmost knot selected by the trend filtering estimator and we define our estimate of the termination of the transit event $\what{T}_1$ as the rightmost knot\footnote{Note that the boundary points of the input space are not considered knots.}. The total transit duration estimate then follows as $\what{T} = \what{T}_1 - \what{T}_0$. Naturally, we define our transit depth estimate $\what{\delta}$ to be the minimum of the relaxed trend filtering estimate.

It is thus far unclear to us whether trend filtering can also reliably detect the end of the ingress and beginning of the egress phases---e.g. via knot selection or examining changes in the estimated derivates---and therefore also provide nonparametric ingress/egress duration measurements. We recommend pairing trend filtering with the traditional analytical transit model search \cite[e.g.,][]{Mandel_2002} for these particular measurements. That is, given the transit depth and total transit duration measurements provided by trend filtering, an analytical planet model can be fit over a parameter space that is constrained by the trend filtering parameter measurements. Coupling the two methods in this way therefore also provides significant computational speedups over a traditional single-stage analytical model search since it greatly reduces the dimensionality of the parameter space to be searched over. Furthermore, the relaxed trend filtering estimate may provide a benchmark $\chi^2$ statistic for the constrained analytical model comparison.

A dedicated paper on this application of trend filtering is forthcoming.

\subsection{Eclipsing binary modeling}
\label{subsec:EB}

\begin{figure*}
\includegraphics[width = \textwidth]{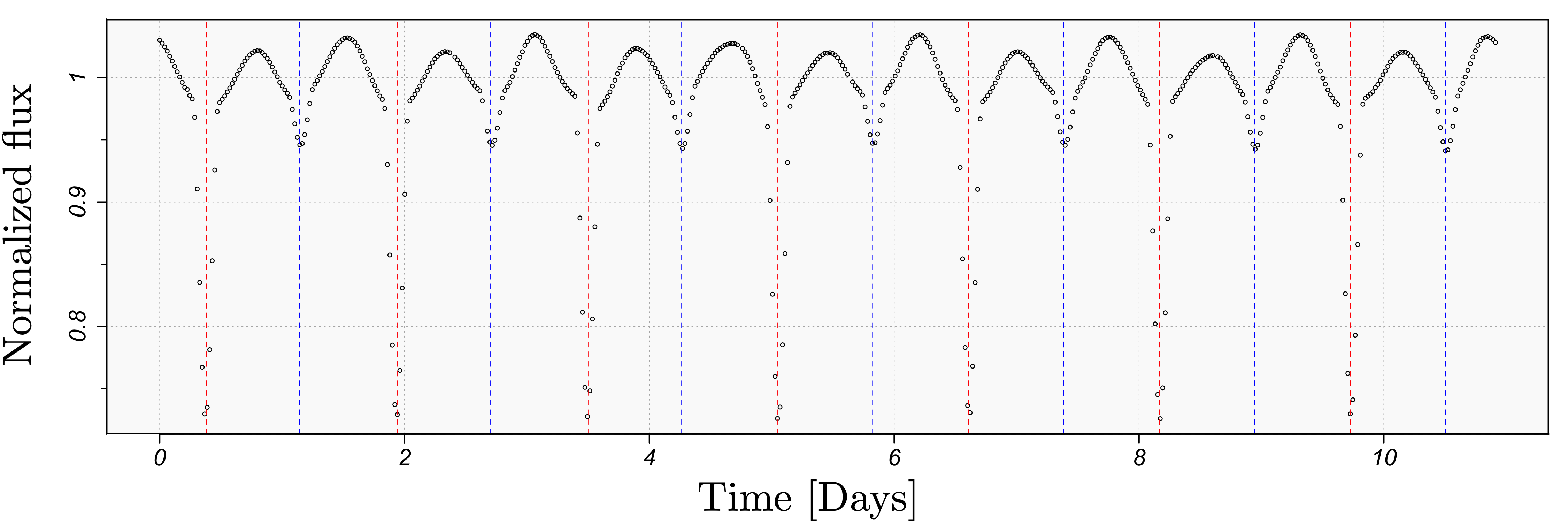}
\vspace{-0.35cm}

\caption{Long-cadence (30-min. increment), detrended, median-normalized light curve of a Kepler eclipsing binary system (KIC 6048106; \citealt{kepler}; \citealt{Prsa_2011}). The vertical red lines mark the primary eclipses (the eclipses of the hotter star) and the vertical blue lines mark the secondary eclipses (the eclipses of the cooler star). KIC 6048106 has an orbital period of $\sim$1.559 days.}
\label{EB_timeseries}
\end{figure*}

In this section we discuss how trend filtering can further improve the work of the Kepler Eclipsing Binary working group, specifically in regard to the Eclipsing Binaries via Artificial Intelligence (EBAI) pipeline used to characterize Kepler eclipsing binary stars via their phase-folded light curves \cite[][]{polyfit,Prsa_2011,Slawson_2011,Matijevi__2012}. The EBAI pipeline utilizes an artificial neural network (ANN) to estimate a set of physical parameters for each binary pair (the temperature ratio, sum of fractional radii, photometric mass ratio, radial and tangential components of the eccentricity, fillout factor, and inclination) from the observables of the phase-folded light curve (e.g., the eclipse widths, depths, and separations). \cite{polyfit} outline the EBAI light curve pre-processing algorithm, which they call \texttt{polyfit}, that provides the crucial step of taking a noisy, irregularly-spaced, phase-folded light curve (detrended for spacecraft motion and normalized by the median flux) and outputting a denoised and gridded phase-folded light curve, which is then fed to the ANN. We propose that trend filtering be used for this pre-processing step instead of the \texttt{polyfit} algorithm for the reasons detailed below.

An eclipsing binary (EB) light curve is characterized by periodic dips in the observed brightness that correspond to eclipse events along the line of sight to an observer. In particular, there are two eclipses per orbital period---a primary and a secondary eclipse. The primary eclipse occurs when the hotter star is eclipsed by the cooler star and produces a comparatively deep dip in observed brightness. Conversely, the secondary eclipse occurs when the cooler star is eclipsed by the hotter star and produces a comparatively shallow dip in the observed brightness. Depending on the effective temperature ratio and orbital period of the EB, the dips may range from very narrow and abrupt to very wide and smooth. In Figure~\ref{EB_timeseries}, we display a detrended, median-normalized, long-cadence (30 min. increment) light curve of a Kepler EB (KIC 6048106; \citealt{kepler}; \citealt{Prsa_2011}), with the primary eclipses and secondary eclipses designated by dashed red and blue lines, respectively. The orbital period of KIC 6048106 is $\sim$1.559 days. 

\begin{figure*}
\includegraphics[width = \textwidth]{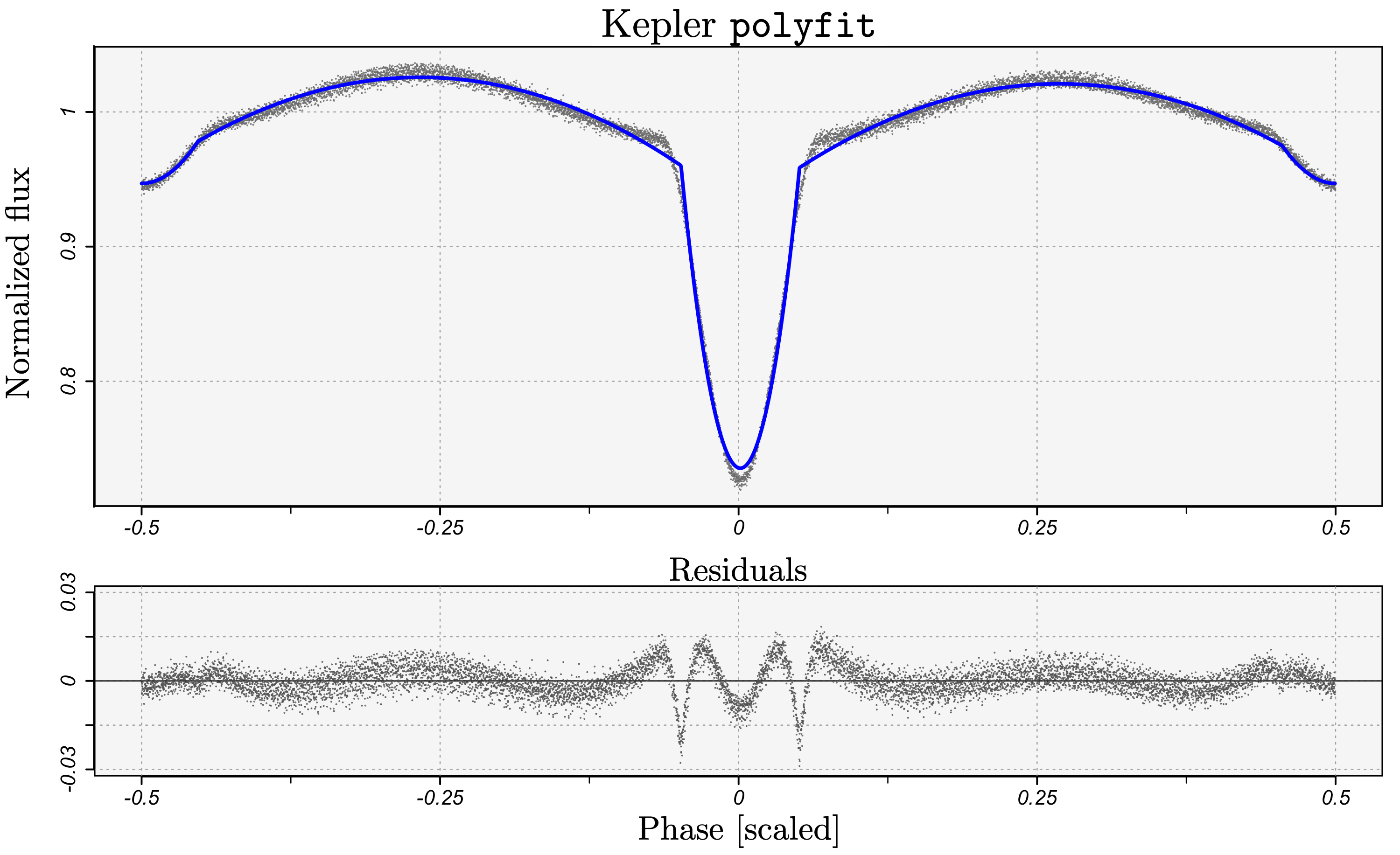} \\
\vspace{0.1cm}

\includegraphics[width = \textwidth]{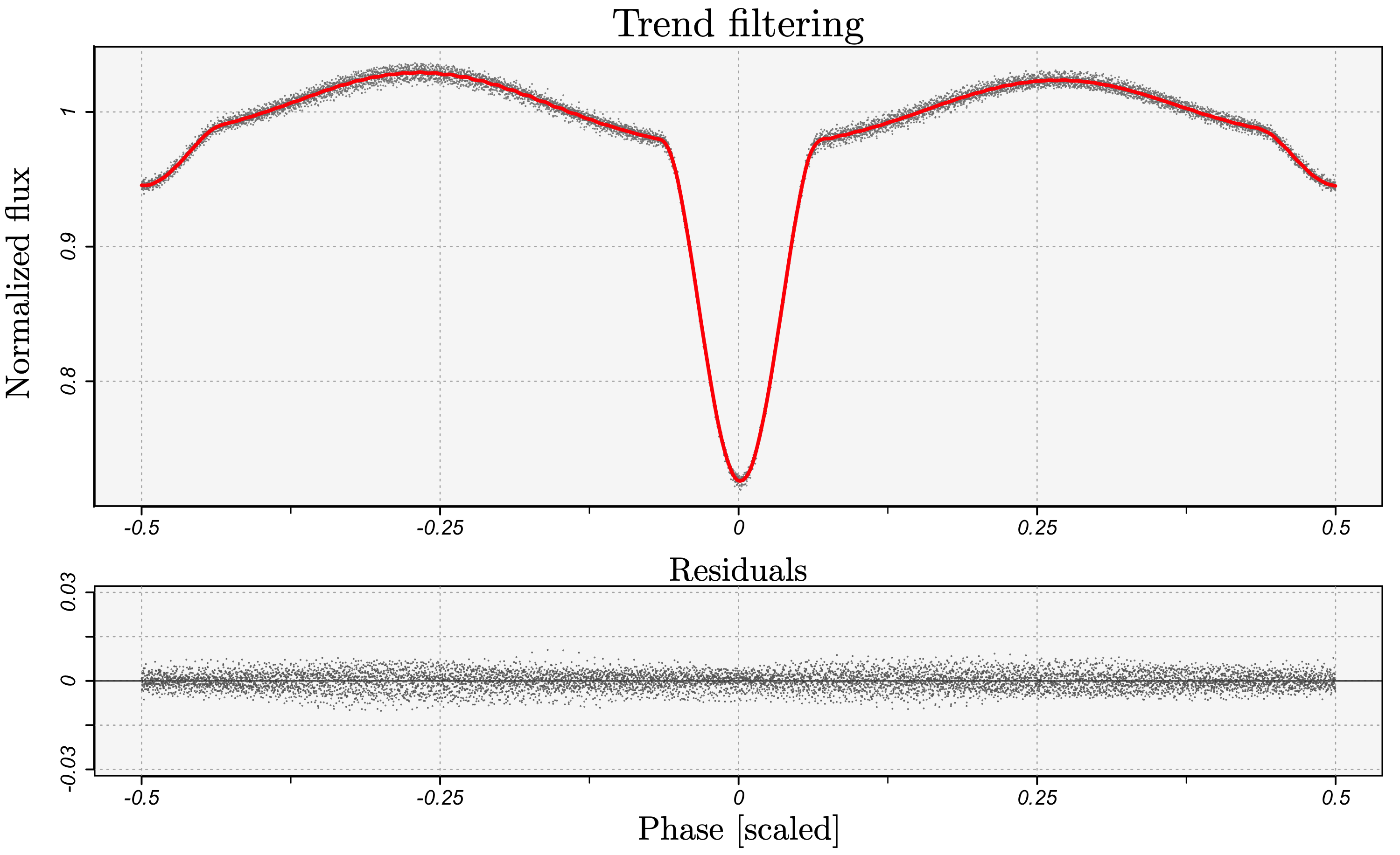}
\vspace{-0.5cm}

\caption{Comparison of the \texttt{polyfit} algorithm of \protect\cite{polyfit} and our trend filtering approach for denoising phase-folded eclipsing binary light curves. The light curve shown in this example comes from the Kepler eclipsing binary system KIC 6048106 (\protect\citealt{kepler}; \protect\citealt{Prsa_2011}). {\bf Top}: The \texttt{polyfit} algorithm fits a piecewise quadratic polynomial by weighted least-squares with four knots selected by a randomized search over the phase space. The estimate is constrained to be continuous but no constraints are enforced on the derivatives at the knots. The overly-stringent assumed model leads to significant statistical bias, which is readily apparent by examining the autocorrelation in the residuals. {\bf Bottom}: Trend filtering is sufficiently flexible to accurately denoise the diverse set of signals observed in phase-folded eclipsing binary light curves. Here, the goodness-of-fit is clear by the random, mean-zero residual scatter.}
\label{EB_phasefolded}
\end{figure*}

After phase-folding with respect to the estimated EB orbital period and centering the primary eclipse at \texttt{Phase~=~0}, the purpose of the EBAI light-curve pre-processing step is to faithfully extract the signal of the phase-folded light curve and evaluate it on a regular grid so that it can then be input into the EBAI ANN. We show a comparison of the \texttt{polyfit} algorithm of \cite{polyfit} and our trend filtering approach in Figure~\ref{EB_phasefolded} on the phase-folded KIC 6048106 light curve. We choose a relatively high S/N light curve here in order to elucidate the significant statistical bias that underlies \texttt{polyfit}. The \texttt{polyfit} algorithm fits a piecewise quadratic polynomial by weighted least-squares with four knots selected by a randomized computational search over the phase space. The piecewise quadratic polynomial is forced to be continuous, but no smoothness constraints are placed on the derivatives of the estimate at the knots. Recalling our discussion in Section~2.1 of Paper~I, this overly-stringent modeling assumption leads to significant statistical bias in the light-curve estimate. The bias is particularly apparent by examining the residuals of the \texttt{polyfit} estimate, which we display below the light curve. Moreover, recalling our discussion of variable-knot regression splines in Section~3.1.1 of Paper~I, a randomized partial search over the space of feasible knots inherently provides no guarantee of finding a global solution---leaving the algorithm susceptible to extreme failure scenarios. We display a quadratic trend filtering estimate of the KIC 6048106 phase-folded light curve in the bottom panel of Figure~\ref{EB_phasefolded}, with the hyperparameter chosen by $K$-fold cross validation. The trend filtering estimate accurately recovers the signal of the light curve (clearly apparent here by examining a high S/N ratio light curve) and produces a desirable random residual scatter about zero. Since the statistical bias introduced by the \texttt{polyfit} pre-processing stage propagates through to the EBAI ANN as systematic bias in the input data, we are confident that the use of trend filtering will in turn improve the error rate of the EBAI ANN output-parameter estimates.

\subsection{Supernova light-curve template generation and estimation of observable parameters}
\label{subsec:supernova}

In this section we demonstrate the use of trend filtering for generating light-curve templates of supernova (SN) events and estimating observable parameters. We illustrate our approach on SN 2016coi (ASASSN-16fp; \citealt{Holoien_2016}) by constructing a B-band light-curve template from the well-sampled observations of \cite{Brown_2014} and \cite{Prentice_2018} and deriving nonparametric estimates and full uncertainty distributions for the maximum apparent brightness, the time of maximum, and the decline rate parameter $\Delta m_{15}(B)$ introduced by \cite{Phillips_1993}. The improvement yielded by trend filtering as a tool for SN light-curve template generation, compared to, for example, the \texttt{SNooPy} cubic smoothing spline approach of the Carnegie Supernova Project \cite[CSP;][]{Contreras_2010,Burns_2010,Burns_2014} primarily corresponds to light curves with an especially bright peak magnitude and fast decline rate. In such cases, trend filtering is better able to recover the abruptness of the peak, the initial sharp decline, and the subsequent slow decay. This behavior is particularly characteristic of Type Ia SNe \cite[e.g.,][]{Woosley_2007}. In cases where the peak is not particularly prominent, trend filtering and cubic smoothing splines produce nearly identical estimates. Our procedure for generating the SN light-curve templates requires reasonably well-sampled observations (in particular, with the initial observation occurring before maximum light). The resulting template libraries can then be used to classify SNe with partially-sampled light curves and derive parameter estimates \cite[e.g.,][]{Belokurov_2004}.

The SN light-curve template generation procedure is analogous to the spectral template generation procedure discussed in Section~\ref{subsec:spectra}, so we discuss it in less formal detail here. Naturally, the same procedure can also be implemented for generating fixed-time spectral templates of SN events. Given a well-sampled light curve, corrected for systematic effects (e.g., $K$-corrections and interstellar reddening), we propose the use of quadratic trend filtering to generate a ``best fit'' to the observations. Given a confident type classification, the trend filtering estimate can then be stored as a light-curve template in the respective library.

\begin{figure*}
\centering
\vspace{-0.1cm}
\includegraphics[width = 0.8\textwidth]{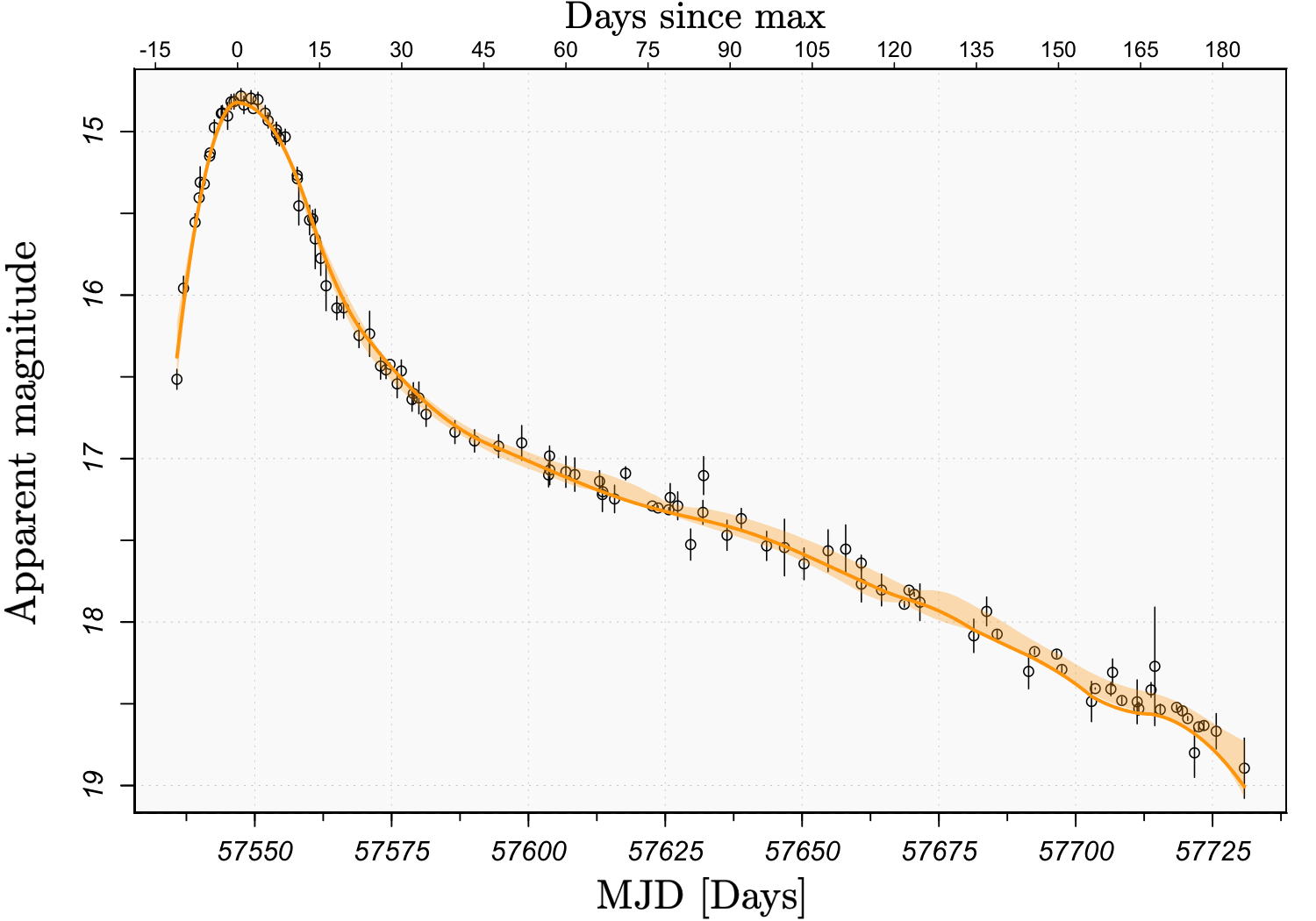} \\
\vspace{-0.4cm}

\hspace{0.01\textwidth}\includegraphics[width = 0.45\textwidth]{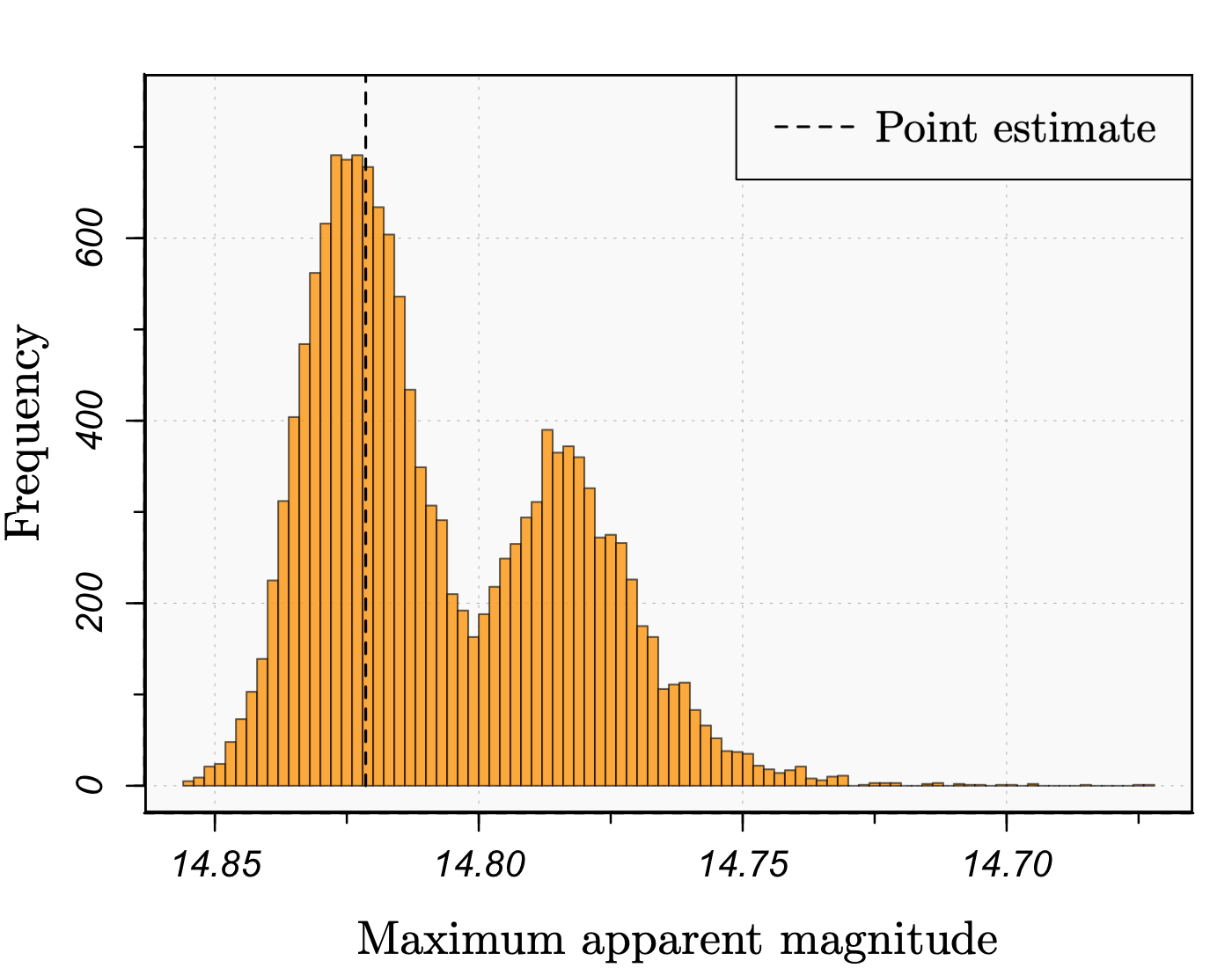} 
\includegraphics[width = 0.45\textwidth]{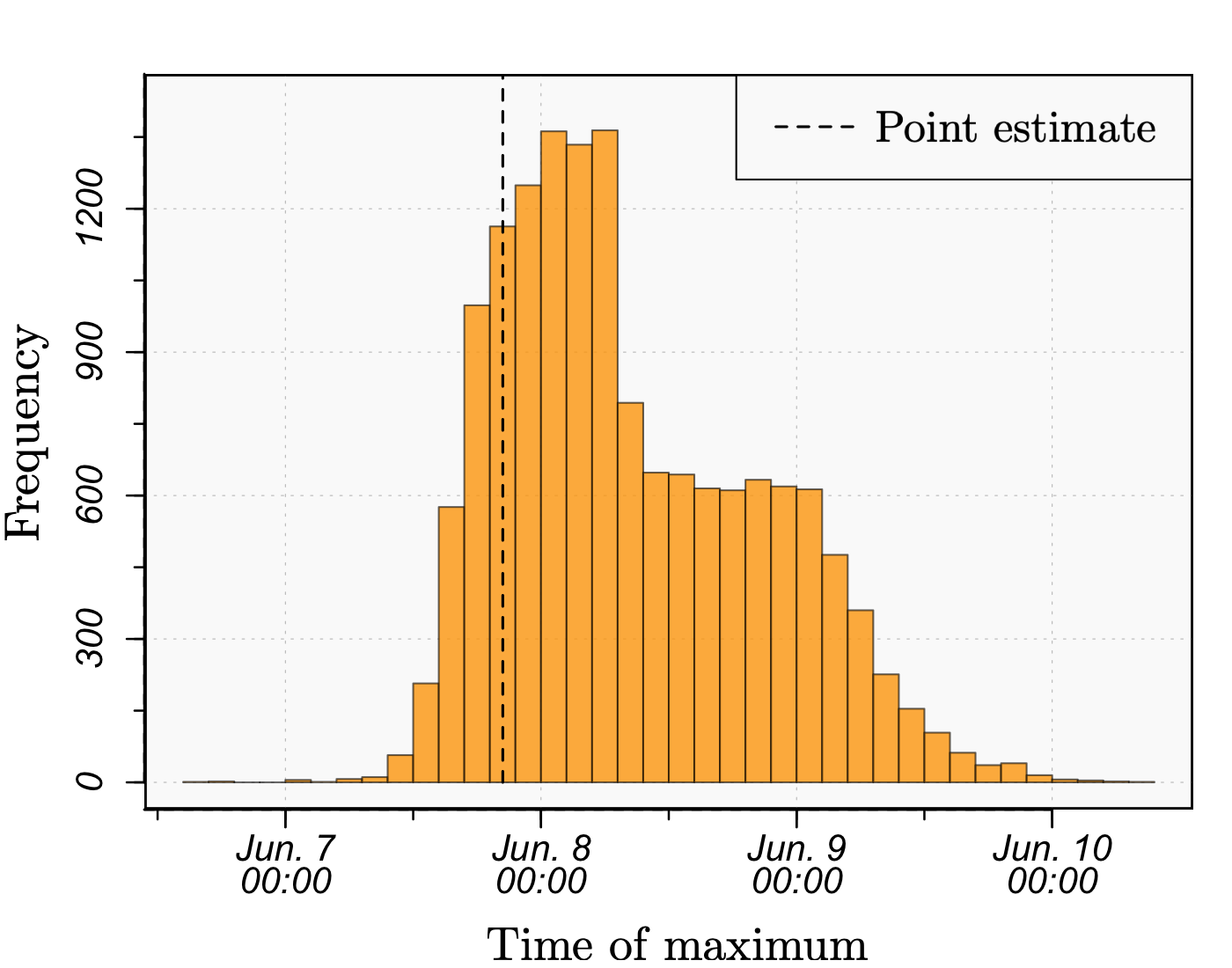} \\
\hspace{0.01\textwidth}\includegraphics[width = 0.45\textwidth]{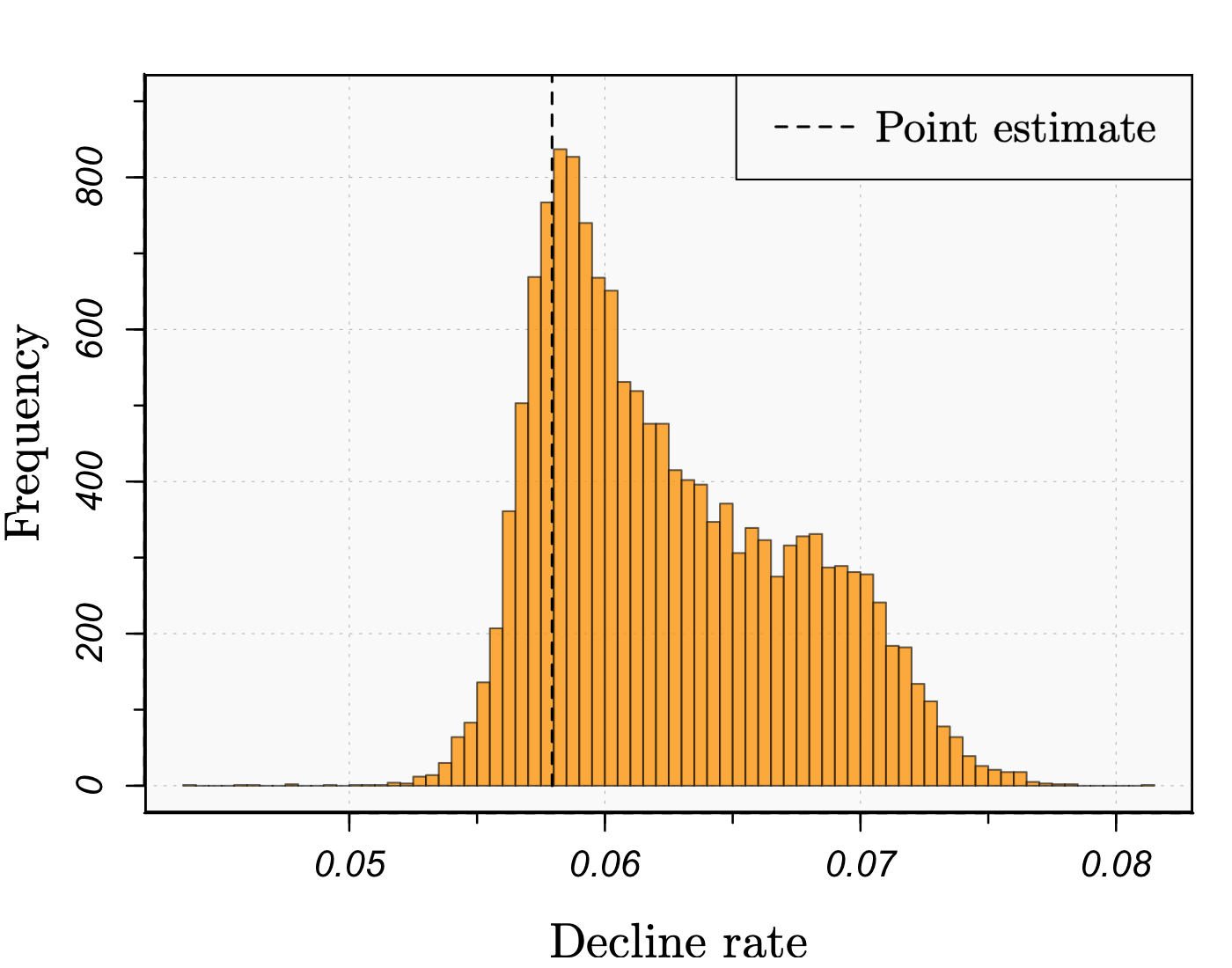} 
\includegraphics[width = 0.45\textwidth]{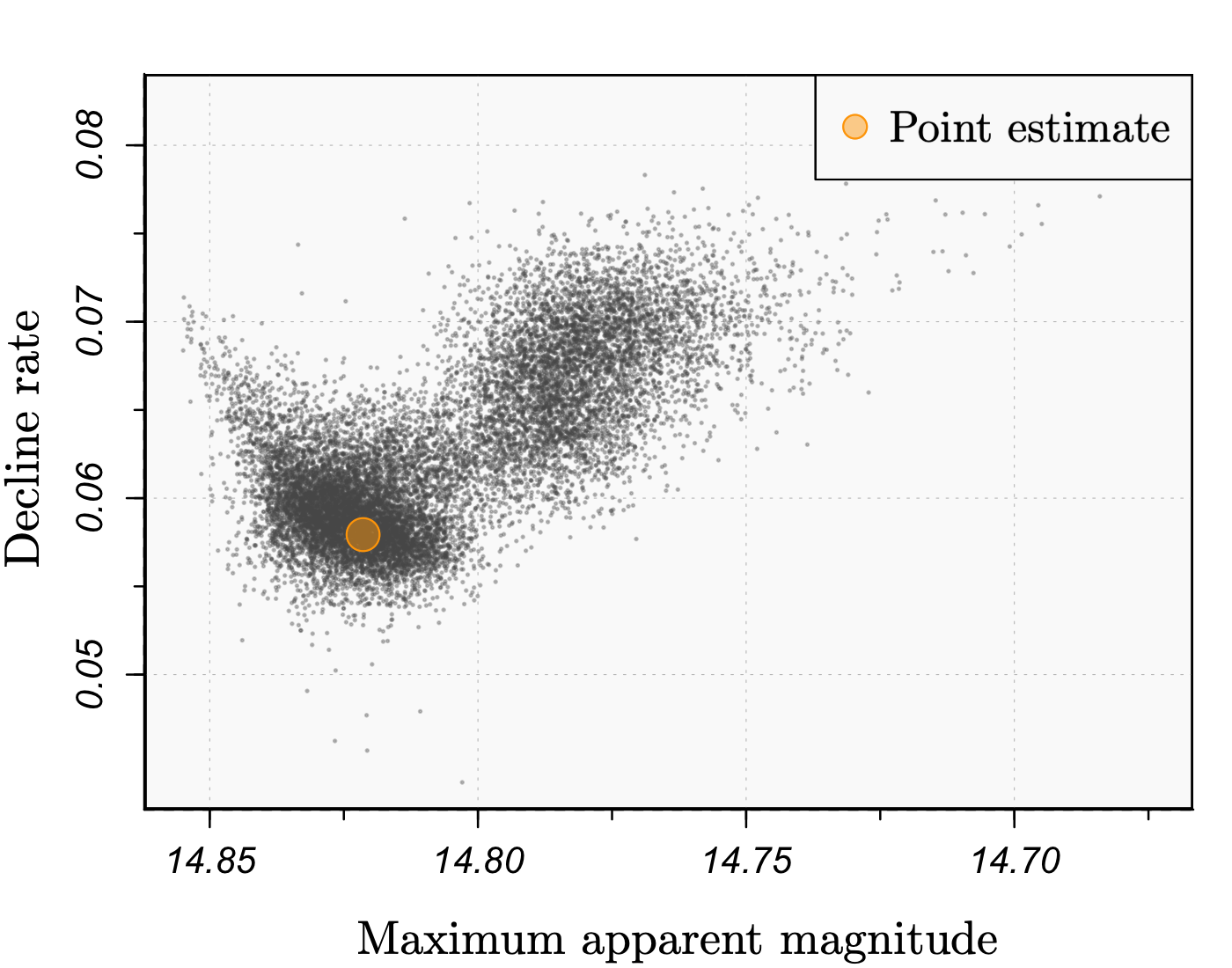} 
\vspace{-0.1cm}

\caption{SN light-curve analysis (SN 2016coi). {\bf Top}: B-band photometry of the supernova SN 2016coi (ASASSN-16 fp; \citealt{Holoien_2016})  discovered on May 27, 2016 by the All Sky Automated Survey for SuperNovae (ASAS-SN; \citealt{Shappee_2014}) in the galaxy UGC 11868 at redshift $z \approx 0.0036$. We fit a quadratic trend filtering estimate, tuned by $K$-fold cross validation, and overlay 95\% nonparametric bootstrap variability bands. {\bf Bottom}: Univariate/bivariate nonparametric bootstrap sampling distributions of the observable parameter estimates derived from the trend filtering light-curve estimate. The bimodality in the bootstrap parameter distributions arises from systematic discrepancies between the observations of the two separate observers \protect\cite[][]{Brown_2014, Prentice_2018}.}
\label{sample_supernova}
\end{figure*}

We show a B-band light curve for SN 2016coi (ASASSN-16fp; \citealt{Holoien_2016}) in the top panel of Figure~\ref{sample_supernova}. The light curve is an aggregation of observations collected by \cite{Brown_2014} and \cite{Prentice_2018}, which we accessed from the Open Supernova Catalog (\protect\citealt{2017ApJ_OSC}). SN 2016coi was discovered on May 27, 2016 (UT 2016-05-27.55) by the All Sky Automated Survey for SuperNovae (ASAS-SN; \citealt{Shappee_2014}) in the galaxy UGC 11868 at redshift $z \approx 0.0036$. The classification of SN 2016coi currently remains intermediate between Type Ib and Type Ic \cite[][]{2016coi_tele,Kumar_2017,Prentice_2018,Terreran_2019}. We fit a quadratic trend filtering estimate to the observations, weighted by measurement uncertainties and with the hyperparameter selected by $K$-fold cross validation. The trend filtering estimate, along with 95\% nonparametric bootstrap variability bands, is overlaid in the top panel of the figure. In the bottom panels we show the univariate nonparametric bootstrap sampling distributions for the estimates of the maximum apparent magnitude, the time of maximum, and the decline rate---defined as the relative change in the B-magnitude light curve from maximum light to the magnitude 15 days after the maximum \cite[][]{Phillips_1993}. We also show a bivariate bootstrap sampling distribution for maximum apparent magnitude versus decline rate. The bimodality in the bootstrap sampling distributions arises from systematic discrepancies between the two separate sets of B-band observations that form the light curve \cite[][]{Brown_2014, Prentice_2018}.

\subsection{Data reduction and compression}
\label{subsec:compression}

Although the primary focus of this paper is the use of trend filtering as a tool for astronomical data analysis, it also possesses potential utility for large-scale reduction and compression of one-dimensional data sets. This owes to several factors: its speed, scalability, flexibility, and representation as a sum of basis functions with a sparse coefficient vector.

Given a one-dimensional set of $n$ observations $(t_1,f(t_1)),\dots,(t_n,f(t_n))\in (a,b)\times\mathbb{R}$, trend filtering can quickly provide a flexible, lower-dimensional approximation of the data where the dimensionality is controlled by the choice of the hyperparameter $\gamma$. In this context $\gamma$ is a subjective choice that specifies the amount of (lossy) compression desired---unrelated to the discussion in Section~3.5 of Paper~I. For any given choice of $\gamma$, let $p$ be the number of knots selected by the trend filtering estimator. The corresponding continuous-time representation of this lower-dimensional approximation is fully encoded by the knot locations and the sparse basis vector with $p+k+1$ nonzero entries, which can be stored efficiently. The falling factorial basis then serves as the ``dictionary'' from which the continuous-time representation can be losslessly recovered. Gridded uncertainty measurements for the reduced observations can also be computed and stored, though not in a sparse format, via the methods discussed in Section~3.6 of Paper~I.

\section{Concluding remarks}

In order to illustrate the broad utility of trend filtering to astronomy, we demonstrated its promising performance on a wide variety of problems across time-domain astronomy and astronomical spectroscopy. We studied the Lyman-$\alpha$ forest of quasar spectra with the most depth---using trend filtering to map the relative distribution of neutral hydrogen in the high redshift intergalactic medium along quasar-observer lines of sight. Furthermore, we discussed how trend filtering can be used to (1) generate galaxy, quasar, and stellar spectral templates and estimate emission-line parameters; (2) produce nonparametric models for exoplanet transit events in phase-folded stellar light curves, providing estimates for the transit depth and total duration; (3) improve upon the \texttt{polyfit} algorithm utilized by the Kepler Eclipsing Binary via Artificial Intelligence (EBAI) pipeline for denoising phase-folded eclipsing binary light curves (as a preliminary step to estimating the physical parameters); (4) generate supernova light-curve templates and produce nonparametric estimates of the maximum apparent magnitude, the time of maximum, and the decline rate; (5) quickly and efficiently compress large collections of one-dimensional data sets. Naturally, we expect trend filtering will find uses in astronomy beyond those that we explicitly discussed.

\section*{ACKNOWLEDGEMENTS}
We gratefully thank Ryan Tibshirani for his inspiration and generous feedback on this topic. This work was partially supported by NASA ATP grant NNX17AK56G, NASA ATP grant 80NSSC18K1015, and NSF grant AST1615940. 

This research contains data collected by the Sloan Digital Sky Survey III (SDSS-III). Funding for SDSS-III has been provided by the Alfred P. Sloan Foundation, the Participating Institutions, the National Science Foundation, and the U.S. Department of Energy Office of Science. The SDSS-III web site is \url{http://www.sdss3.org/}. SDSS-III is managed by the Astrophysical Research Consortium for the Participating Institutions of the SDSS-III Collaboration including the University of Arizona, the Brazilian Participation Group, Brookhaven National Laboratory, Carnegie Mellon University, University of Florida, the French Participation Group, the German Participation Group, Harvard University, the Instituto de Astrofisica de Canarias, the Michigan State/Notre Dame/JINA Participation Group, Johns Hopkins University, Lawrence Berkeley National Laboratory, Max Planck Institute for Astrophysics, Max Planck Institute for Extraterrestrial Physics, New Mexico State University, New York University, Ohio State University, Pennsylvania State University, University of Portsmouth, Princeton University, the Spanish Participation Group, University of Tokyo, University of Utah, Vanderbilt University, University of Virginia, University of Washington, and Yale University. This research has made use of the NASA Exoplanet Archive, which is operated by the California Institute of Technology, under contract with the National Aeronautics and Space Administration under the Exoplanet Exploration Program. This paper includes data collected by the Kepler mission. Funding for the Kepler mission is provided by the NASA Science Mission Directorate. This research contains data that was accessed from the Open Supernova Catalog, a centralized, open repository for supernova metadata, light curves, and spectra. The Open Supernova Catalog web site is \url{https://sne.space}. The original sources of the supernova light curve studied in this paper are credited in the text.




\bibliographystyle{mnras}
\addcontentsline{toc}{section}{Bibliography}
\bibliography{mybib}{}
\bibliographystyle{mnras}
\setcitestyle{authoryear,open={[},close={]}}

\appendix

\section*{Appendix}

\subsection*{Mock quasar Lyman-$\alpha$ forest reduction}
\label{A1}

The \cite{Bautista} mock quasar catalog is designed to mimic the observational data generating processes of the quasar spectra released in Data Release 11 of the Baryon Oscillation Spectroscopic Survey (\citealt{Alam}). We pool the first three realizations of the mock catalog, i.e. \texttt{M3\_0\_3/000}, \texttt{M3\_0\_3/001}, and \texttt{M3\_0\_3/002} and remove all damped Ly$\alpha$ systems (DLAs), Lyman-limit systems (LLS), and broad absorption line quasars (BALs). We assume no metal absorption in the \lya forest and correct estimation and subtraction of the sky. We mask all pixels with \texttt{and\_mask} \texttt{$\neq$} \texttt{0} or \texttt{or\_mask} \texttt{$\neq$} \texttt{0}. Finally, we retain only the spectra with $\geq500$ pixels in the truncated \lya forest and those with a fixed-input-optimal trend filtering hyperparameter satisfying $\gamma_0 < 5.25$. Spectra with $\gamma_0 \geq 5.25$ correspond to the very lowest S/N ratio observations, where the trend filtering estimate typically reduces to a global power law fit (zero knots). The final mock sample contains 124,709 quasar spectra.

\bsp	
\label{lastpage}
\end{document}